\documentclass[aps,prl,reprint,superscriptaddress,longbibliography]{revtex4-2}

\usepackage{graphicx} 
\usepackage{dcolumn} 
\usepackage{bm} 
\usepackage{hyperref}
\usepackage{physics}
\usepackage[dvipsnames]{xcolor}

\begin{document}

\title{Probing the hollowing transition of a shell-shaped BEC with collective excitation}


\author{Zerong Huang}
\thanks{These two authors contributed equally to this work.}
\affiliation{Department of Physics, The Chinese University of Hong Kong, Hong Kong SAR, China}
\author{Kai Yuen Lee}
\thanks{These two authors contributed equally to this work.}
\affiliation{Department of Physics, The Chinese University of Hong Kong, Hong Kong SAR, China}
\author{Chun Kit Wong}
\thanks{Current address: Institut f\"ur Experimentalphysik, Universit\"at Innsbruck, Austria.}
\affiliation{Department of Physics, The Chinese University of Hong Kong, Hong Kong SAR, China}
\author{Liyuan Qiu}
\affiliation{Department of Physics, The Chinese University of Hong Kong, Hong Kong SAR, China}
\author{Bo Yang}
\thanks{Current address: Xi'an Institute of Applied Optics, Xi'an, China.}
\affiliation{Department of Physics, The Chinese University of Hong Kong, Hong Kong SAR, China}
\author{Yangqian Yan}
\affiliation{Department of Physics, The Chinese University of Hong Kong, Hong Kong SAR, China}
\affiliation{State Key Laboratory of Quantum Information Technologies and Materials, The Chinese University of Hong Kong, Hong Kong SAR, China}
\author{Dajun Wang}
\email{djwang@cuhk.edu.hk}
\affiliation{Department of Physics, The Chinese University of Hong Kong, Hong Kong SAR, China}
\affiliation{State Key Laboratory of Quantum Information Technologies and Materials, The Chinese University of Hong Kong, Hong Kong SAR, China}

\date{\today}

\begin{abstract}

We investigate the hollowing transition of a shell-shaped Bose-Einstein condensate using collective excitations. The shell is created using an immiscible dual-species BEC mixture, with its hollowness controlled by tuning the repulsive interspecies interaction via a Feshbach resonance. Our results reveal two distinct monopole modes in which the two condensates oscillate either in-phase or out-of-phase. The spectrum of the out-of-phase mode exhibits a non-monotonic dependence on the interspecies interaction, providing a clear signature of the topology change from a filled to a hollow condensate. Furthermore, we find that the critical point of the hollowing transition depends strongly on the number ratio of the two species. Our findings provide a detailed understanding of the topology change in shell-shaped quantum gases and pave the way for future study of quantum many-body phenomena in curved spaces.

\end{abstract}

\maketitle

\textit{Introduction---} The study of double-species Bose-Einstein condensates (BECs) dates back to 1957~\cite{Khalatnikov1957}, when the superfluid helium mixture was theoretically studied for the first time. Following the creation of BECs of dilute atomic gases in 1995, interest in this topic was revived, leading to intense theoretical and experimental explorations still ongoing to date. Although earlier studies primarily focused on phase separation, that is, the miscible-immiscible phase transition for repulsive intraspecies and interspecies interactions~\cite{Ho1996,Esry1997,Law1997,Pu1998a,Pu1998b,Timmermans1998,Sinatra1999,Hall1998,Papp2008,Modugno2002,McCarron2011,Lercher2011,Wacker2015,Wang2015,Maddaloni2000,Cavicchioli2022}, the recent discovery of the quantum liquid droplet phase in the mean field collapsing regime~\cite{Petrov2015,Cabrera2018,Semeghini2018,Errico2019,Guo2021} suggests that there is still a wealth of physics to explore in the dual-species BEC system.

Creating and investigating shell-shaped BECs based on immiscible dual-species BEC systems is another newly established research direction in this field~\cite{Jia2022,Wolf2022,Tononi2024}. The shell topology bestows BECs with distinctive features such as periodic boundaries, local curvature,  and two surfaces, which are absent in BECs in standard bulk geometries. These features can lead to a variety of unique properties, including the emergence of self-interference during free expansion~\cite{Lannert2007,Tononi2020,Jia2022}, and the formation of vortex and anti-vortex pairs under fast rotation~\cite{Padavic2020}. Two decades after the initial proposal, shell BECs have only recently been successfully produced using several different methods, after overcoming the distortion of shell potentials caused by gravity~\cite{Carollo2022,Jia2022}. The method based on immiscible double BECs allows the production of shell BECs without the need for a microgravity environment, making it more convenient for further exploration. For instance, the self-interference phenomenon was already studied in the first experiment based on this method~\cite{Jia2022}.

In this work, we study another interesting feature of the shell BEC: its hollowing transition, using double species $^{23}$Na and $^{87}$Rb BECs with tunable interspecies interactions. We employ the monopole mode of collective excitation as an indicator of the hollowness of the shell BEC~\cite{Wolf2022,Padavic2017,Sun2018}, which is controlled by a Feshbach resonance between the $^{23}$Na and $^{87}$Rb atoms. In the context of the dual-species BEC system, this study is also directly related to collective excitation across the miscible-immiscible phase transition, which has been theoretically studied previously~\cite{Pu1998b,Wolf2022}, but has not been investigated experimentally. We note that the miscible to immiscible phase separation transition in dual-species BECs has been previously studied by several groups by observing morphological changes of the condensates~\cite{Thalhammer2008,Papp2008,McCarron2011,Wang2015}. Yet, this approach often fails to identify a distinct transition point due to its limited sensitivity and the confounding effects of the trapping potential. Here, in the out-of-phase monopole mode between the $^{23}$Na and $^{87}$Rb BECs with increasing repulsive interspecies interaction strengths, we observe a clear critical point as the $^{23}$Na BEC transitions to a shell shape with a hollow center. We also take this as the starting point of the dual-species BEC phase separation transition. In addition, we also find that this point depends sensitively on the atom number ratios between the two condensates.

\begin{figure}[t]
\begin{center}
\includegraphics[width =0.85\linewidth]{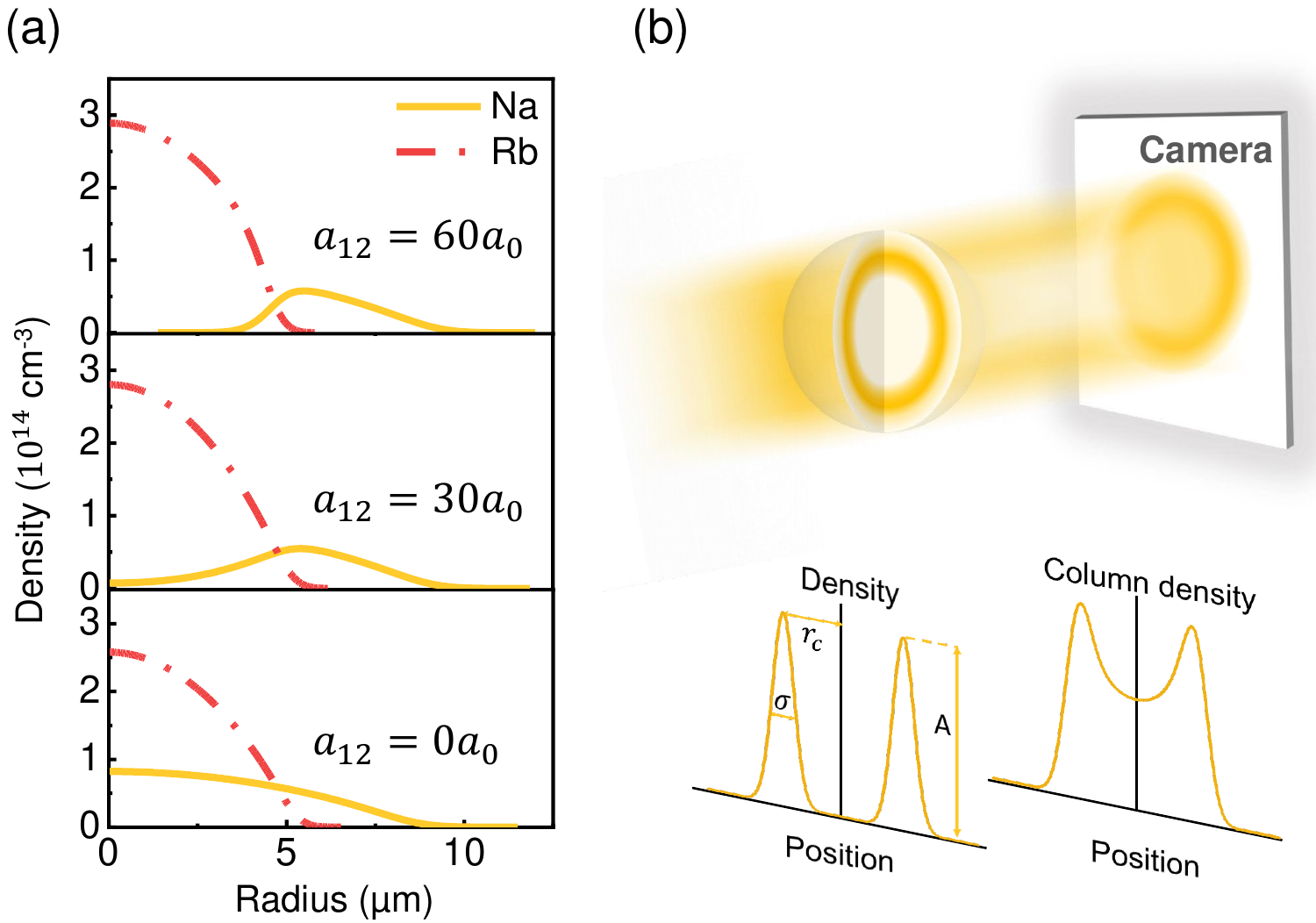}
\end{center}
\caption {Creating and probing a shell BEC in a $^{23}$Na-$^{87}$Rb double species BEC system. (a) From bottom to top: calculated density distributions of the two condensates with increasing interspecies scattering length $a_{12}$ illustrate the miscible-immiscible phase transition and the hollowing transition of the Na shell. (b) During absorption imaging, a hollow shell appears as a double-peaked feature. The bottom sub-figures depict the central intersections of the original shell (left) and its projection along the probing beam direction (right). The size of the shell can be extracted from the projected distribution using our fitting protocol. }
\label{fig1}
\end{figure}

\textit{Experiment---} Our experiment starts with a dual-species BEC of ${}^{23}$Na and ${}^{87}$Rb atoms co-trapped in an optical dipole trap formed by crossing three orthogonally propagating 946 nm laser beams. To simplify the collective excitation spectrum, we create a nearly spherical harmonic potential by carefully adjusting the power ratios between the three laser beams. The measured oscillation frequencies of the trap along different axes are consistent with each other to within 5\%. At the 946 nm ``magic'' wavelength~\cite{Safronova2006,Jia2022}, the two species experience the same trap oscillation frequency $\omega_0$ and thus the same gravitational sag $-g/\omega_0^2$ in the vertical direction. This ensures the centers of mass of the two condensates nearly overlap.

We prepare both $^{23}$Na and $^{87}$Rb atoms in their lowest hyperfine Zeeman level $\ket{F=1,m_F=1}$. Away from Feshbach resonances, the interaction constants satisfy $g_{12}\geq \sqrt{g_{11} g_{22}}$, rendering the two condensates immiscible. Here, $g_{ij} = 2\pi \hbar^2 a_{ij}/\mu_{ij}$ where $a_{ij}$ are the $s$-wave scattering lengths, $\mu_{ij} = m_i m_j/(m_i + m_j)$ are the reduced masses, and $m_i$ are the atomic masses, respectively (with $i,j = 1$ for $^{23}$Na and 2 for $^{87}$Rb). Under these conditions, the $^{23}$Na BEC will form a shell surrounding the $^{87}$Rb BEC~\cite{Jia2022}. 

To control the hollowness of the $^{23}$Na BEC, we use a magnetic Feshbach resonance at $B_0 = 347.65$ G to tune $a_{12}$ following $a_{12} = a_{bg} (1+\Delta/(B-B_0))$. Here $a_{bg} = 76.3a_0$ is the background $^{23}$Na-$^{87}$Rb scattering length near $B_0$, and $\Delta = 4.26$ G is the width of the resonance~\cite{Guo2022}. By adjusting the magnetic field $B$ from 351.91 G to 370 G, we can modify $a_{12}$ from $0a_0$ to $60a_0$ while keeping $a_{11}=60.5a_0$~\cite{Knoop2011} and $a_{22} = 100.14 a_0$ \cite{Kempen2002} constant. As shown in Fig.~\ref{fig1}(a), numerical simulations using the coupled Gross-Pitaevskii equations (GPEs) suggest that within the range of $a_{12}$, intermediate state between a bulk and a shell sample of $^{23}$Na BEC can be achieved.

\begin{figure}[t]
    \centering
    \includegraphics[width =0.85\linewidth]{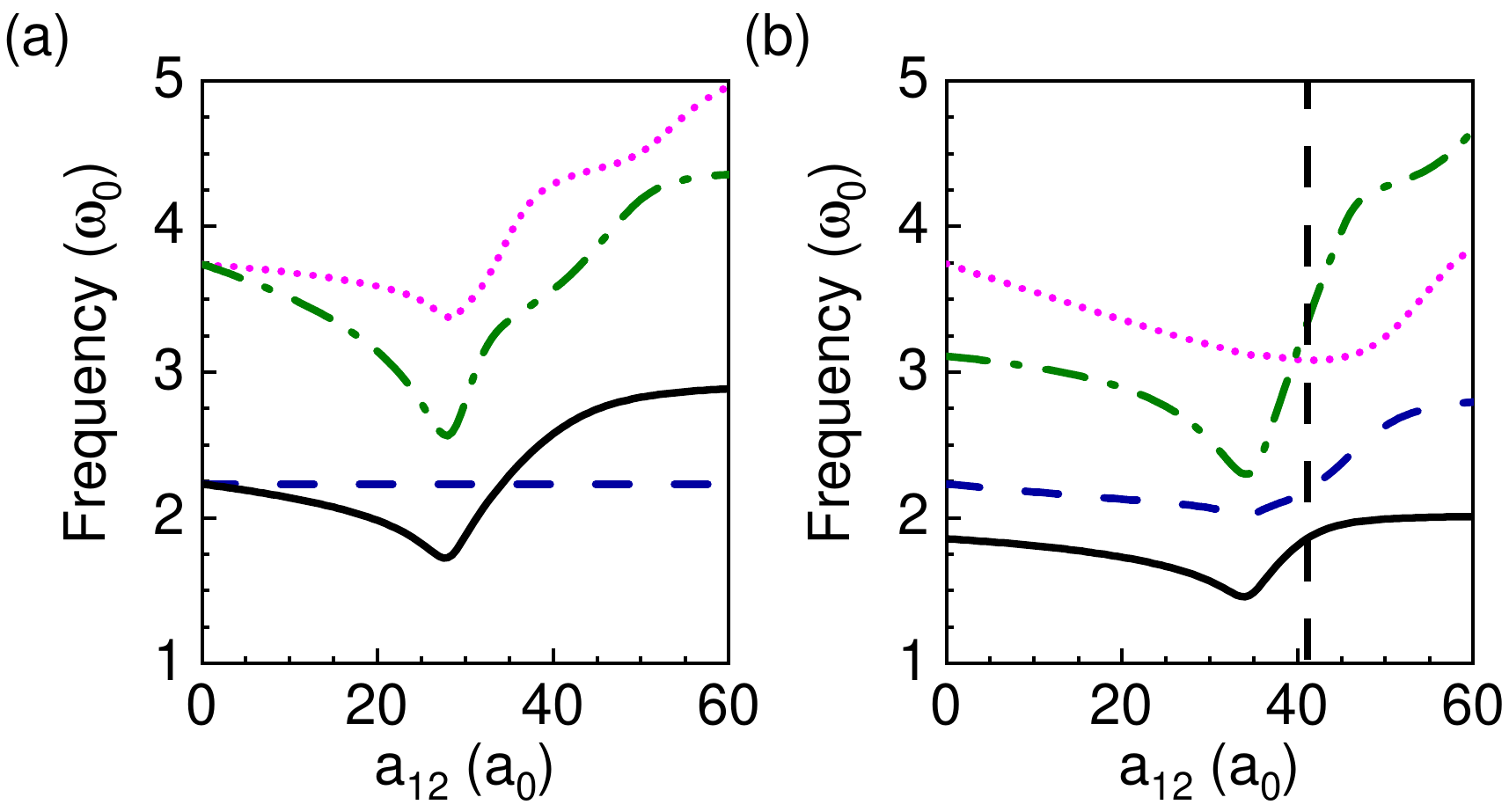}
    \caption{ Simplifying the double BEC excitation spectrum with magic wavelength spherical potential. All plots are calculated using $N_{1}=N_{2}=10^6$. 
    (a) and (b) show the lowest four modes in magic-wavelength and non-magic-wavelength spherical traps, respectively. For the former, the trap oscillation frequencies are $2\pi\times 118.6~\text{Hz}$ for both species, while for the latter, it is $2\pi\times 98.6~\text{Hz}$ for $^{87}$Rb. While in (a) the lowest in-phase (blue dashed curve) and out-of-phase modes (black solid curve) have no coupling, the same modes in (b) are coupled together as evidenced by the avoided crossing and the gap at the position marked by the vertical dashed line.  
    }
    \label{fig2}
\end{figure}

\textit{In-phase and out-of-phase monopole modes---} 
Analogous to classical coupled oscillators, the collective excitation of dual-species BECs also includes in-phase and out-of-phase modes. In general, the two modes are coupled~\cite{Pu1998b, Wolf2022}, probing them independently in experiments presents significant challenges. However, this issue can be mitigated by using the magic-wavelength spherical trap. Figure~\ref{fig2}(a) and (b) show the numerically calculated spectra as a function of the interspecies scattering length for several of the lowest monopole modes using Bogoliubov-de Gennes equations (BdGEs), with the trapping light frequency set at the magic condition and slightly deviated from it, respectively. In the magic wavelength case, the two modes are fully decoupled and exhibit a real crossing, thus allowing them to be probed with minimal ambiguity. Conversely, in the latter case, the coupling leads to an avoided crossing with an energy gap, resulting in a switch between the in-phase and out-of-phase modes. The coupling appears even for small non-zero trap frequency difference $\Delta \omega$ and the gap moves to different $a_{12}$ when $\Delta \omega$ is tuned~\cite{Note1}. Furthermore, we define and calculate a quantity called the two-species collectivity, which equals one when both species contribute equally to the excitation and approaches zero when one species dominates. For $\Delta \omega = 0$ in the magic-wavelength trap, the collectivity remains high even for small $a_{12}$. However, it rapidly decreases for non-zero $\Delta \omega$, indicating that the excitation loses its two-species nature~\cite{Note1}. Thus, to clearly distinguish the in- and out-of-phase modes, especially at small $a_{12}$, the magic-wavelength spherical trap is essential. In addition, a spherical potential supports monopole modes without damping~\cite{Lobser2015} and is generally easier to handle with analytical and computational methods. These advantages make the magic-wavelength spherical trap an ideal choice for detailed comparisons between our measurements and theoretical models.

We use two very different modulation methods to excite these two modes. In the first experiment, we excite and study the in-phase monopole mode by modulating the trapping potential. We first prepare the dual-species BEC at a target interspecies scattering length $a_{12}$ by tuning the magnetic field. Subsequently, we modulate the power of the three trapping beams with the same phase and amplitude, which induces synchronized compression and expansion of the two condensates. The trap modulation amplitude and duration need to be set carefully to maximize the amplitude of the monopole oscillation while avoiding the excitation of other collective modes. Empirically, we determine that a modulation amplitude of 4\% and a duration of approximately 15 modulation periods can induce large enough monopole oscillation amplitude without significantly coupling to other modes. 

To excite the out-of-phase mode, we instead vary $a_{12}$ by applying a sinusoidal modulation to the magnetic field while keeping the trapping potential constant. This modulation induces anti-phased changes to the sizes of the two condensates. For instance, when $a_{12}$ is increased, the $^{23}$Na cloud is forced outward, causing it to expand, while the $^{87}$Rb cloud is compressed inward, resulting in a size decrease. The dynamic interplay between the two condensates under this modulation scheme leads to the out-of-phase monopole oscillations of our interest. 

Similar to that for the in-phase mode, we empirically choose the amplitude and duration of the $a_{12}$ modulation to obtain the maximum possible signal without significant excitation of other collective modes. For small $a_{12}$, we use a moderate modulation amplitude of $2a_0$ which is enough to cause significant sample size variations. However, when $a_{12}$ becomes large enough to cause phase separation, we increase the modulation amplitude to $5a_0$ to compensate for the reduced overlap and effectively excite the desired oscillation.

After modulation, we allow the two condensates to evolve in the trap for varying durations. Finally, we release them from the trap and image the resulting clouds using two-species high-magnetic-field absorption imaging method after 15 ms of free expansion~\cite{Jia2020}. As illustrated in Fig.~\ref{fig1}(b), to obtain the size of the $^{23}$Na shell, we model it with a three-dimensional spherical Gaussian shell function $n_1 \times \mathrm{exp}(-(r-r_c)^2/\sigma^2)$. We fit the absorption images using the Abel transformation of this Gaussian shell function to extract the shell center $r_c$, shell thickness $\sigma$, and peak density $n_1$. We use $r_c$ to represent the size of the $^{23}$Na shell. For the $^{87}$Rb BEC size and $^{23}$Na bulk sample when $a_{12}$ is small, we use the average of the horizontal and vertical rms widths obtained from 2D Gaussian fits.

\begin{figure}[t]
    \centering
    \includegraphics[width =0.85\linewidth]{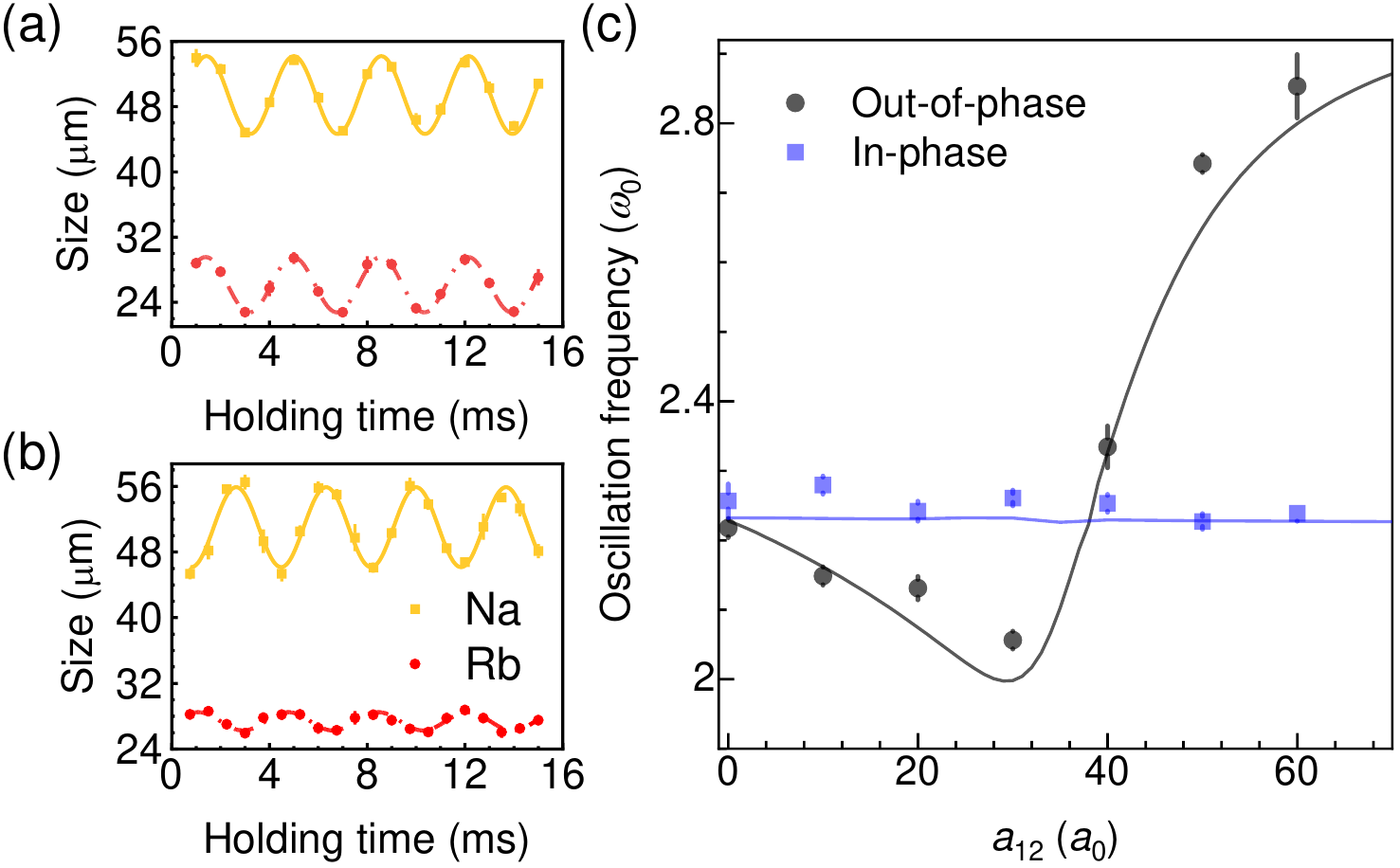}
    \caption{(a) In-phase size oscillation for $^{87}$Rb and $^{23}$Na at $a_{12}=30a_0$. For $^{87}$Rb, data points represent the averaged horizontal and vertical sizes extracted from 2D Gaussian fittings of the images, while for $^{23}$Na, $r_c$ obtained from the shell fitting procedure are used. (b) The out-of-phase size oscillation at $a_{12}=30a_0$. (c) Frequency spectrum for the lowest in-phase and out-of-phase modes. The blue and black solid lines are the calculations from BdGEs for the lowest in-phase and out-of-phase monopole modes, respectively. For this set of measurements, the $^{23}$Na atom number is $1.0(2)\times10^5$ and the $^{87}$Rb atom number is $7.0(5)\times10^4$. All oscillation frequencies $\omega$ are normalized to the trap frequency $\omega_0$. The error bars of $\omega$ are from the sinusoidal fitting. }
    \label{fig3}
\end{figure}

Figure~\ref{fig3} (a) and (b) are exemplary resulting monopole oscillations of two miscible condensates with $a_{12} = 30 a_0$ excited by modulating the trapping potential and the interspecies interaction strength, respectively. For the former case, the measured phase slip between the $^{23}$Na and $^{87}$Rb size oscillations is less than $0.1 \pi$, which is consistent with in-phase oscillation; for the latter in Fig.~\ref{fig3}(b), this is about $1.1 \pi$, i.e., the two condensates oscillate out-of-phase with each other. As expected for a spherical potential~\cite{Lobser2015}, the damping is minimal during the observation period. The slight phase slip and damping are attributed to residual mixing between the two modes, most possibly due to the imperfect spherical symmetry and unavoidable anharmonicity of the optical potential.

\textit{The hollowing transition---} To investigate the hollowing transition, we examine the monopole modes for $a_{12}$ from 0 to around 60$a_0$. As will be discussed later, the oscillation frequency $\omega$ of the out-of-phase mode also depends on the atom numbers. The atom number fluctuations in our system are large enough to generate observable effects. To mitigate this problem, we used only post-selected data points with atom number fluctuations within 20\% for this measurement.

The blue points in Fig.~\ref{fig3}(c) show the measured oscillation frequency $\omega$ of the in-phase mode, which barely changes with $a_{12}$. This is reminiscent of the in-phase normal mode of two classical coupled oscillators with the same natural frequencies, where the coupled oscillation frequency is the same as that of the individual uncoupled oscillators and is not affected by the coupling. Here, the measured $\omega$ is $\sqrt{5}\omega_0$, exactly the same as that of the monopole mode for individual BECs in the Thomas-Fermi (TF) regime~\cite{pethickdilutegas}. This agrees with our theoretical derivation~\cite{Note1} which shows that the two-species BEC can be effectively treated as a single one in this mode. Obviously, this mode is not sensitive to the hollowing transition.


The behavior of the out-of-phase mode is drastically different, as shown by the black points in Fig.~\ref{fig3}(c). For two non-interacting condensates at $a_{12} = 0a_0$, the oscillation frequency $\omega$ is also $\sqrt{5} \omega_0$, the same as that of the in-phase mode. As $a_{12}$ increases, $\omega$ first decreases to a minimum of approximately $2\omega_0$ at $30a_0$. Afterwards, it starts to increase and eventually levels off for $a_{12} \geq 60a_0$ when the Na shell is fully formed. This non-monotonic dependence on $a_{12}$ thus makes this mode a sensitive probe of the hollowing transition.

While this behavior agrees fully with our numerical solution based on BdGEs [black solid curve in Fig.~\ref{fig3}(c)], an intuitive understanding can be gained from the fact that the out-of-phase mode involves density oscillations transverse to the condensate boundaries, where the relative motion between the two species makes $\omega$ sensitive to the overlap, and thus $a_{12}$~\cite{Padavic2017}. In addition, this mode predominantly excites the $^{23}$Na shell, while the bulk $^{87}$Rb BEC is driven to respond with an opposite-sign motion to minimize the interaction energy. This is evident from the $\pi$-phase difference between the two species and the larger oscillation amplitude of $^{23}$Na, as shown in Fig.~\ref{fig3}(b). This suggests that we can gain insight by studying the thin-shell limit with $N_1\ll N_2$~\cite{Lannert2007}, where $^{23}$Na dominates the mode dynamics and $^{87}$Rb excitation, being the response, becomes less important. Here $N_1$ and $N_2$ are the numbers of $^{23}$Na and $^{87}$Rb atoms, respectively.  This allows an analysis with the simplified BdGEs, which can quantitatively reproduce the frequency spectrum of the full BdGEs~\cite{Note1}. 


Under such a limit, the $^{87}$Rb BEC merely acts as a background, contributing to an effective potential 
$V_{\text{eff}}(r)=\frac{1}{2}m_{1}\omega_0^2 r^2+g_{12}n_{2}(r) $
for $^{23}$Na. Here $n_2(r)$ is the ground-state density distribution of $^{87}$Rb. For the small $a_{12}$ region before the shell starts to form, since the two condensates are miscible, under the TF approximation, 
\begin{equation}
V_{\text{eff}}(r) = \frac{1}{2}m_1\tilde{\omega}_0^2r^2 + C, 
\end{equation}
where 
\begin{equation}
\tilde{\omega}_0 = \omega_0 \left(1 - \frac{g_{12} m_2}{2 g_{22} m_1}\right) 
\end{equation}
is a weakened trap frequency and $C$ is a constant shift. The simplified two-species BdGEs reduces the system to the single-species case with a collective oscillation frequency $\omega =\sqrt{5}\tilde{\omega}_0$~\cite{Note1}. For cases with more balanced numbers $N_{1}\sim N_{2}$, we can use a hydrodynamic analysis instead~\cite{Note1}, which gives $\omega = \sqrt{5}\omega_0\left(1-\frac{g_{12}m_{1}}{g_{11}m_{2}}\right)$. For both scenarios, the reduction of $\omega$ with increasing $a_{12}$ before the hollowing transition is well accounted for by these analyses.


The post-hollowing increase in $\omega$ can also be understood with the effective potential $V_{\text{eff}}(r)$. At large $a_{12}$, when the inner surface of the shell is formed, the shell experiences a skewed ``V"-shape $V_{\text{eff}}(r)$, with its minimum at the equilibrium position $r_c$. Approximating $V_{\text{eff}}(r)$ as harmonic, its steepness qualitatively determines $\omega$ of the shell BEC. As $a_{12}$ increases, the potential becomes steeper and $\omega$ increases. However, at very large $a_{12}$, $V_{\text{eff}}(r)$ transforms into a hard wall potential plus a linear term. Further strengthening of the wall no longer affects the dynamics, leading to the observed plateau in the frequency spectrum.


From a physical standpoint, at small $a_{12}$, the ${}^{23}$Na BEC oscillates like an accordion \cite{Lannert2007}, with only its width changing. In this regime, density modulations are localized in the center, where both condensates experience a weakened trap.
This leads to reduced densities and lower stiffness, resulting in lower $\omega$ that does not depend on $N_1$. 
As $a_{12}$ increases, the $^{87}$Rb condensate becomes a rigid core for the $^{23}$Na shell, creating an inner boundary.
This boundary restricts the motion of the width and therefore restores the stiffness. Furthermore, it liberates and shifts the dominant motion degree of freedom to $r_c$, where the ${}^{23}$Na BEC oscillates like a balloon \cite{Lannert2007} with higher $\omega$.
Thus, the emergence of the inner boundary changes the trend, creating a minimum in the out-of-phase mode frequency spectrum. 
This minimum is a hallmark of the transition point.



\begin{figure}[t]
    \centering
    \includegraphics[width=0.95\linewidth]{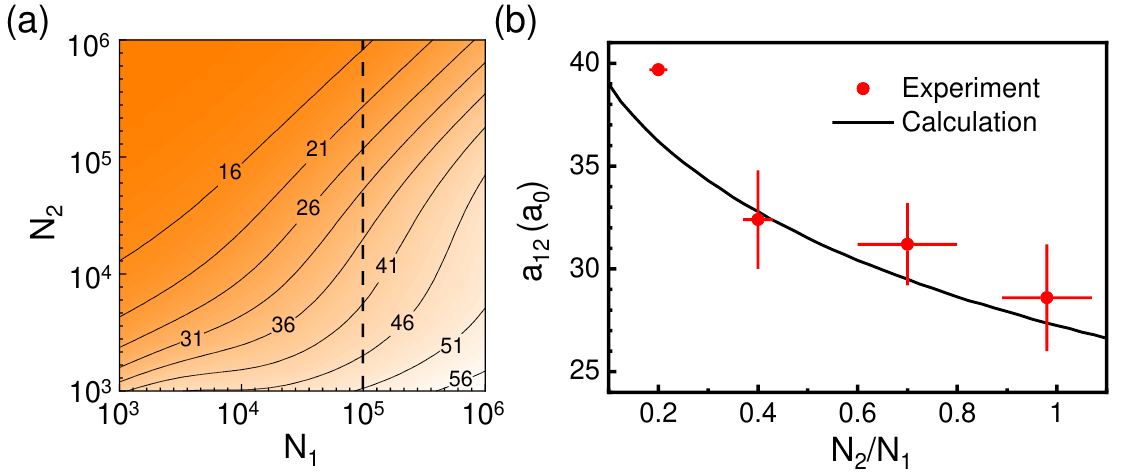}
    \caption{The effect of number ratio on the hollowing transition point. (a) The numerically calculated critical $a_{12}$ (in units of $a_0$) for the hollowing transition with different combinations of $^{23}$Na and $^{87}$Rb atom numbers. (b) Experimentally measured critical $a_{12}$ for several number ratios $N_2/N_1$ between $^{87}$Rb and $^{23}$Na. The $^{23}$Na number is fixed at approximately $1\times 10^5$, thus the theoretical curve corresponds to the red vertical bar in (a). The error bars of $N_2/N_1$ are from statistics of number fluctuations, while those of the critical $a_{12}$ are from the fitting.}
    \label{fig4}
\end{figure}

\textit{The effect of atom numbers---} Next, we investigate the dependence of the hollowing transition on the number of atoms. To this end, we measure the out-of-phase oscillation frequency $\omega$ as a function of $a_{12}$, similar to that shown in Fig.~\ref{fig3}(c), for various ratios of $^{87}$Rb to $^{23}$Na atom numbers $N_{2}/N_{1}$. Figure~\ref{fig4}(a) shows the calculated general behavior of the critical $a_{12}$ for all combinations of atom numbers ranging from $10^3$ to $10^6$, illustrating that the hollowing transition point is highly sensitive to the atom number ratios. In the experiment, we fix $N_{1}$ at approximately $10^5$ and vary $N_{2}$ from $2 \times 10^4$ to $10^5$ for each set of measurements. This allows us to probe the out-of-phase monopole mode for $N_{2}/N_{1}$ ranging from 0.2 to 1. We then empirically fit each spectrum with a bi-Gaussian function, sharing the same center positions but having different widths, to extract the critical $a_{12}$ at the minimum of $\omega$, which corresponds to the onset of the hollowing transition.

The measured critical $a_{12}$ as a function of number ratio $N_{2}/N_{1}$ is summarized in Fig.~\ref{fig4}(b). The observed decrease in the critical $a_{12}$ with increasing $N_{2}/N_{1}$ indicates that the $^{23}$Na BEC forms a shell structure at progressively smaller $a_{12}$ values as the $^{87}$Rb number increases. This behavior is attributed to the contribution of the $^{87}$Rb BEC to $V_{\text{eff}}(r)$. As the repulsion experienced by $^{23}$Na from $^{87}$Rb is $g_{12}n_2 \sim g_{12} N_2^{2/5}$, increasing $N_2$ hardens the $^{87}$Rb core and lowers the interaction strength needed for $^{23}$Na to become hollow. 

However, it is worth noting that when the number of atoms in either species becomes very low, the quantum pressure term starts to play a significant role in the hollowing transition. This explains why the critical $a_{12}$ saturates in the upper-left and lower-right regions of Fig.~\ref{fig4}(a). The larger deviation of the data point at $N_2/N_1=0.2$ from the theoretical curve can be attributed to the less pronounced frequency minimum, which reduces the reliability of the fitting procedure used to extract it. 



\textit{Conclusion---}
 In summary, we have identified two distinct oscillation modes, in-phase and out-of-phase, in our shell BEC system. The in-phase mode frequency remained constant during the transition to a hollow shell, while the out-of-phase mode displayed a non-monotonic dependence on the interspecies scattering length, signaling the topological change to a shell geometry. We further utilized the unique frequency dip property of the out-of-phase mode to investigate the impact of the number ratio of the two species on the hollowing transition point. We found that our shell BEC hollows out more readily at larger interspecies scattering lengths when the number ratio of $^{87}$Rb to $^{23}$Na is smaller. Our results reveal yet another unique feature of the shell BEC and also probe the double BEC miscible to immiscible phase transition with unprecedented resolution.

\begin{acknowledgments}
This work was supported by the Hong Kong RGC General Research Fund (Grants 14301620 and 14302722), and the Collaborative Research Fund C4050-23GF.
\end{acknowledgments}

\end{document}


\preprint{}

\title{Supplementary material: Probing the hollowing transition of a shell-shaped BEC with collective excitation}

\date{\today}



\maketitle



\section{Experimental methods and data analysis}

\subsection{The spherical magic wavelength trap}

\begin{figure}[h]
    \centering
    \includegraphics[width=0.5\linewidth]{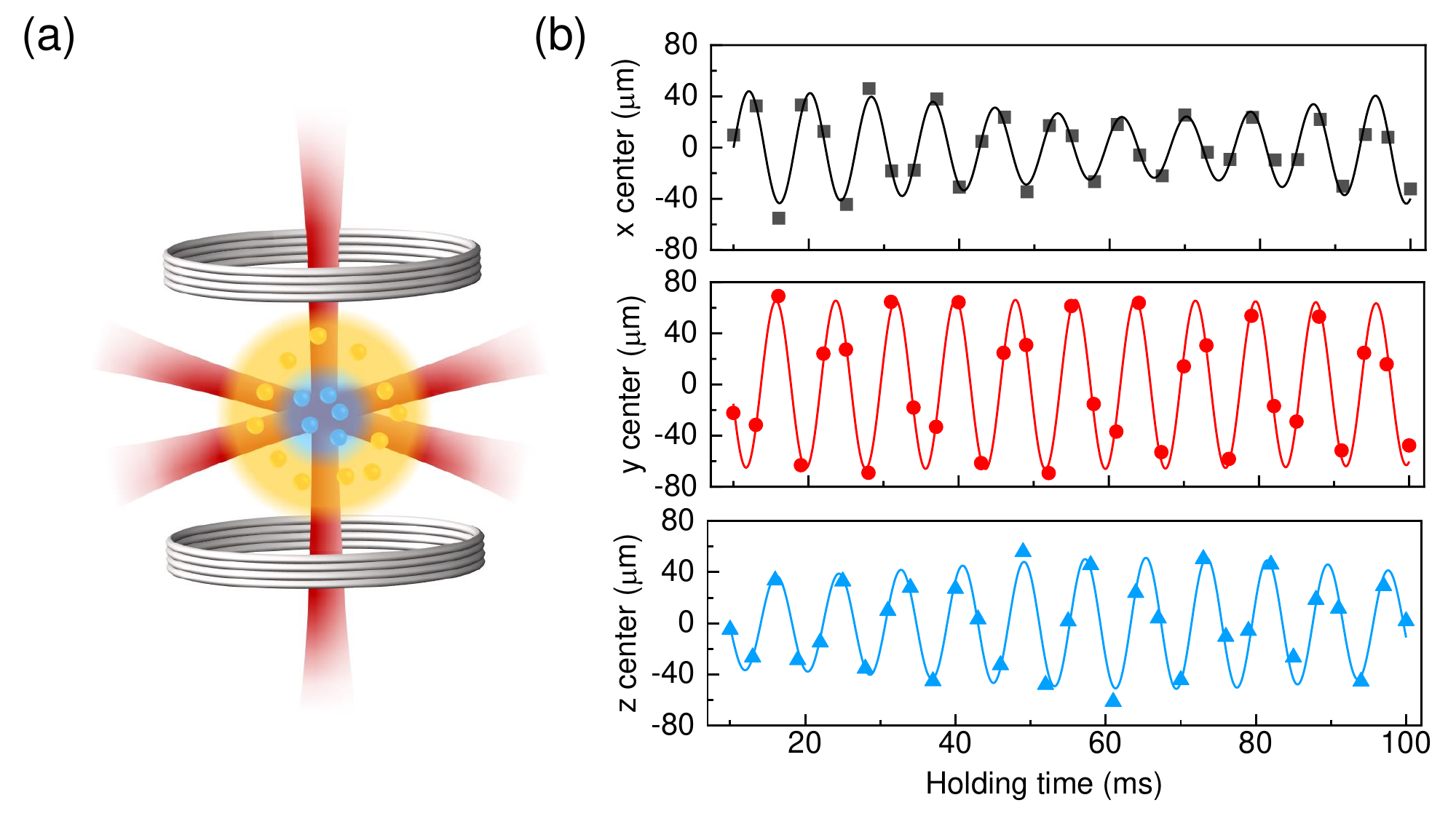}
    \caption{(a) Creating a magic wavelength spherical optical potential by crossing three laser beams. The asphericity is fined tuned by adjusting the relatively powers of the beams. (b) Characterizing the spherical trap via sloshing motions simultaneously in all three directions. For this set of data, the trap frequencies obtained from the global fitting are $(\omega_{x'},\omega_{y'},\omega_{z'}) = 2\pi\times [122.3(3),128.6(7),124.8(1)]$ Hz. (see text for details)}
    \label{sFig1}
\end{figure}

The procedure for preparing the shell BEC sample is similar to our previous work~\cite{Wang2015,Jia2022}. Initially, we evaporate the double BEC in a crossed 1070 nm optical dipole trap and subsequently load them into a 946 nm optical dipole trap at low magnetic field. We then ramp up the magnetic field to reach the desired value. To realize a spherical potential, we then introduce an additional 946 nm laser beam in the vertical direction [Fig.~\ref{sFig1}(a)]. By fine tuning the power of this vertical beam, we finally obtain equal trap frequencies along all directions.  

The trap frequency of the spherical trap is determined by observing the sloshing motion of the BEC sample from two different directions~\cite{Scherer2013,Lobser2015}. We begin by applying a small magnetic gradient pulse for a short duration to displace the sample from its equilibrium position and then turn off the gradient. After a variable evolution time, we release the BEC from the trap and image it after 15 ms of time-of-flight. The axes of the trapping potential are defined in a rotated frame ($x', y', z'$) relative to the reference frame ($x, y, z$) set by the imaging axis. In the frame of the trapping potential, the position of the atomic cloud is given by:
\begin{equation}
     \mathbf{r}' = \begin{bmatrix}
A_{x'} \sin\left( \omega_{x'} t + \phi_{x'} \right) \\
A_{y'} \sin\left( \omega_{y'} t + \phi_{y'} \right) \\
A_{z'} \sin\left( \omega_{z'} t + \phi_{z'} \right)
\end{bmatrix}
\end{equation}
 where $A_{i'}$, $\omega_{i'}$, and $\phi_{i'}$ ($i' = x', y', z'$)  are the amplitude, angular frequency, and phase of the oscillation along the eigenaxes of the potential. In the imaging axis frame, the position of the atomic cloud is given by: 
 \begin{equation}
      \mathbf{r} = \mathbf{R}_z(\theta_z) \mathbf{R}_y(\theta_y) \mathbf{R}_x(\theta_x) \mathbf{r}'
    \label{eqposition}
 \end{equation}
 where $R_i(\theta_i)$ is the rotation matrix of the which rotates the coordinate system around axis $i'$.
 
We perform a global fit of the Eq.~(\ref{eqposition}) to the atomic cloud position obteained from both horizontal and vertical imaging. An example of the observed oscillation is shown in Fig.~\ref{sFig1}(b). The geometric mean of the trap frequencies  $\omega_0 = \left( \omega_{x'} \omega_{y'} \omega_{z'} \right)^{1/3}$  is used for data analysis. 
 
Since the spherical potential is critical for decoupling the in-phase and out-of-phase modes, and for minimize the effect of damping, we typically collect data using a spherical potential with a residual asphericity less than 5\%. The asphericity is defined as $(\omega_{max} - \omega_{min}) / \omega_0$, where $\omega_{max}$ and $\omega_{min}$ are the maximum and minimum of the three trap frequencies, respectively.

\subsection{Shell Fitting}
\begin{figure}[h]
    \centering
    \includegraphics[width=0.3\linewidth]{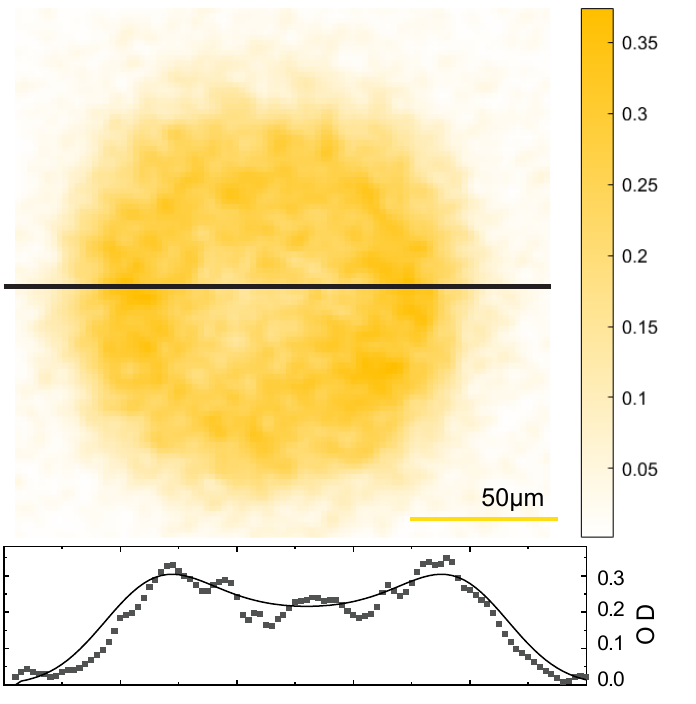}
    \caption{Example of the fitting results. The bottom panel shows the fitting results for the data points indicated by the black line in the upper image.}
    \label{fig:fitting_example}
\end{figure}

Instead of using the lightsheet imaging method as in our previous work~\cite{Jia2022}, we adopt a shell fitting method to better extract information from the shell~\cite{Sanz2022}. An example of the fitting is shown in Fig. \ref{fig:fitting_example}.

\subsection{Transition Point}

To extract the critical interspecies scattering length $a_c$ for the hollowing transition, we fit the data empirically using a bi-Gaussian function with the same center position but different widths
\begin{equation} 
\label{Bigaussian}
f(x) = 
\begin{cases}
A \exp\left(-\dfrac{(a_{12} - a_c)^2}{2\sigma_L^2}\right) + B, & a_{12} < a_c, \\[6pt]
A \exp\left(-\dfrac{(a_{12} - a_c)^2}{2\sigma_R^2}\right) + B, & a_{12} \ge a_c.
\end{cases} 
\end{equation} 
Here, $A$ is the amplitude, $B$ is the offset, and $\sigma_L$ and $\sigma_R$ are the widths on the left and right sides of $a_c$, respectively.

\begin{figure}[h]
\centering 
\includegraphics[width=0.3\linewidth]{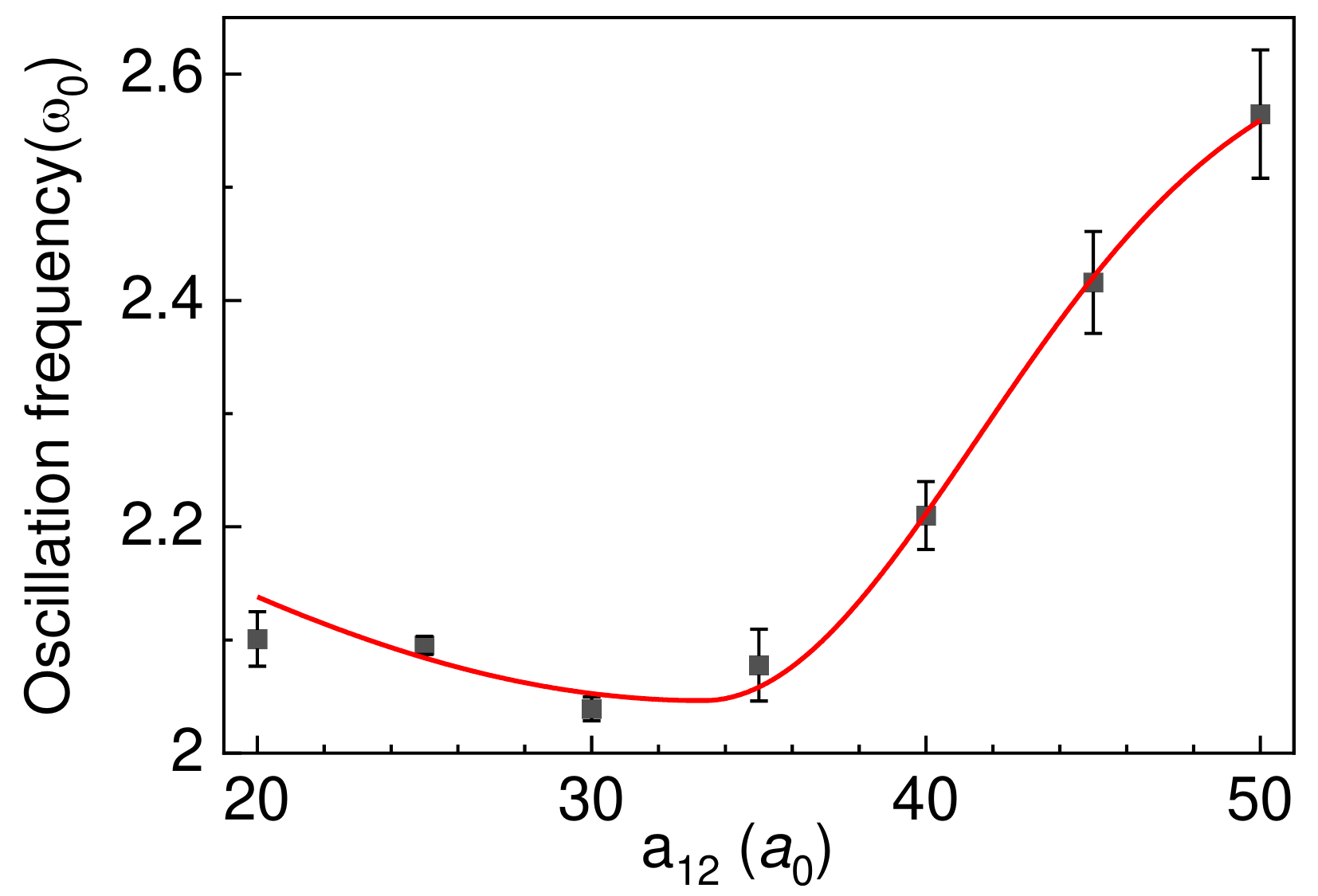} 
\caption{Bi-Gaussian fit of the out-of-phase mode spectrum for extracting the transition point of the hollowing transition. The black squares are the measured oscillation frequencies at different $a_{12}$, while the curve is from the fitting to the Bi-Gaussian in Eq.~\ref{Bigaussian}.} 
\label{sFig.3} 
\end{figure}

An example is shown in Fig. \ref{sFig.3}, where for $N_2/N_1=0.4$ the critical interspecies scattering length of the hollowing transition is determined to be  $a_c = 32(2)a_0$ from the fitting.

\section{Theoretical analysis}

\subsection{Collectivity}
 
We define collectivity between two species as
\begin{equation}
    \text{Collectivity} = 1 - \tanh\left(2 \cdot \frac{\abs{\Delta N_{1} - \Delta N_{2}}}{\Delta N_{1} + \Delta N_{2}}  \right),
\end{equation}
where $\Delta N_i = \int \abs{\delta n_i(r)}  d^3r$ is the number of particles involved in the excitation with density modulation $\delta n_i(r)$. The factor $2$ serves as a scale where the case $\Delta N_i = 3\Delta N_j$ defines large particle imbalances, and the hyperbolic tangent enhances the sensitivity to imbalances.

\begin{figure}[h]
    \centering
    \includegraphics[width =0.4\linewidth]{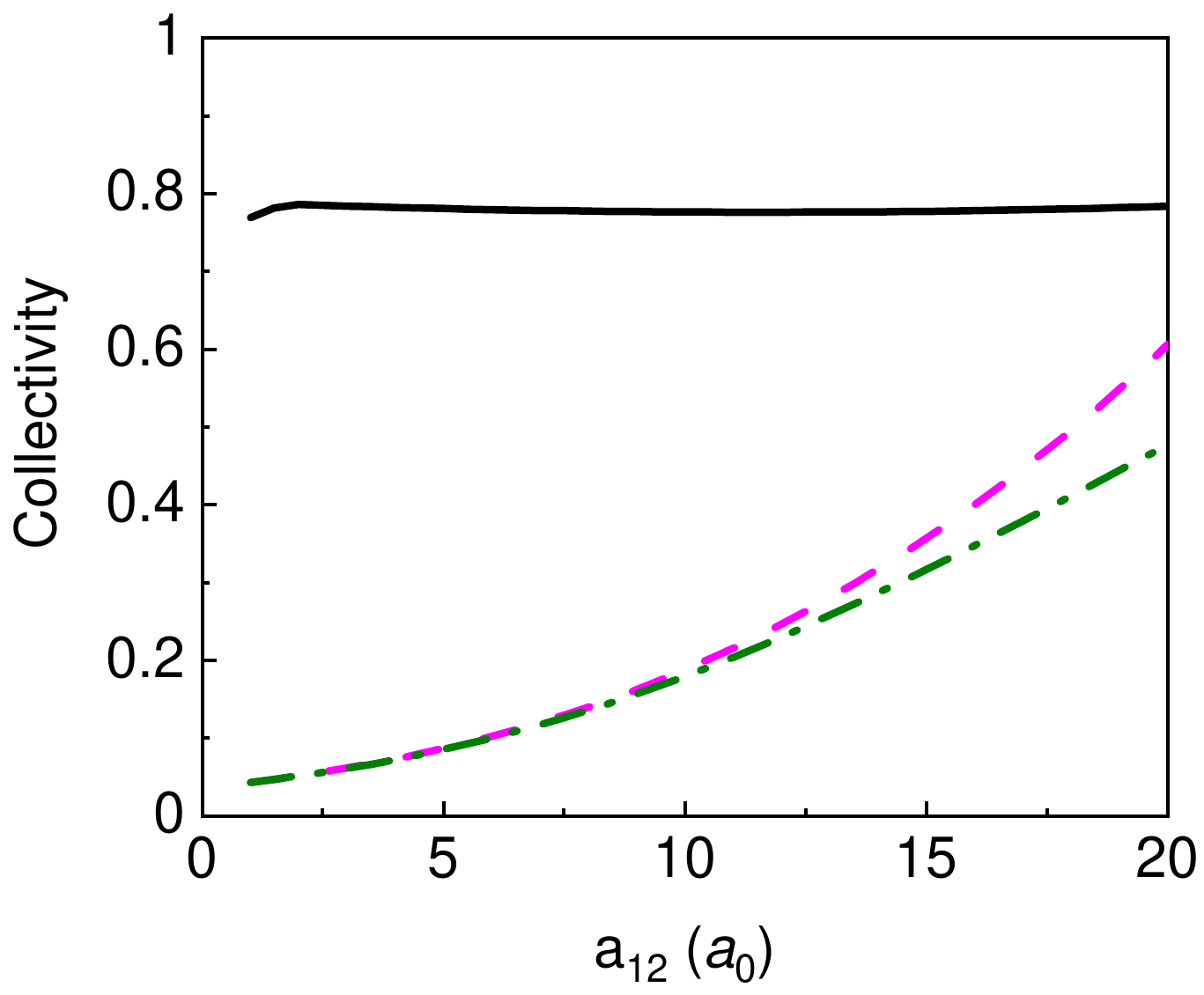}
    \caption{Collectivity between species excitations, showing the average of in-phase and out-of-phase modes. Solid line: trap frequencies $f_{1}=f_{2}=118.6\text{Hz}$; dashed: $f_{1}=108.6\text{Hz}$, $f_{2}=118.6\text{Hz}$; dash-dotted: $f_{1}=118.6\text{Hz}$, $f_{2}=108.6\text{Hz}$. Data at $a_{12}=0a_0$ is omitted. For large $a_{12}$ , the collectivity is always high due to the interspecies interaction. 
    }
    \label{collectivity}
\end{figure}

With this definition, the collectivity is 1 when both species contribute equally to the excitation and approaches 0 when one species dominates. As shown in Fig.~\ref{collectivity}, this is particularly relevant for small $a_{12}$ when the interaction is weak. At matched trapping frequencies, the collectivity remains high even in this regime. However, when the trap frequencies are different for the two species, the collectivity rapidly decreases, indicating that the excitations lose their two-species nature. In this sense, the magic wavelength spherical trap is necessary in order to probe the hollowing transition. For larger $a_{12}$, collectivity is always high as interaction already couples two species.


\subsection{Hydrodynamics Equation Method for Balanced $N_1 \sim N_2$}

\subsubsection{Single-Species Hydrodynamics Equations}
From the Thomas-Fermi (TF) approximation, the ground-state density distribution for the single species case in an isotropic harmonic trap is \cite{Pethick2008}
\begin{equation}
    n_i(r)=\frac{1}{2}\frac{m_i}{g_{ii}}\omega_0^2(R_i^2-r^2).
    \label{SingleSpeciesTF}
\end{equation}
$R_i$ is determined from the conservation of the number of particles.
To consider the collective excitation, a perturbative term is added such that the general density distribution is written as $n_i(\mathbf{r},t)\mapsto n_i(r)+\delta n_i(\mathbf{r})e^{-\mathbf{i}\omega t}$, this lead to an eigenvalue problem \cite{Pethick2008}
\begin{equation}
    -m_i\omega^2\delta n_i(\mathbf{r})=\nabla\cdot[g_{ii}n_i(r)\nabla\delta n_i(\mathbf{r})].
    \label{SingleSpeciesMain}
\end{equation}
We are interested in the breathing mode with $\omega=\sqrt{5}\omega_0$ and \cite{Pethick2008}
\begin{equation}
    \delta n_i(\mathbf{r})=C_i\left(1-\frac{5}{3}\frac{r^2}{R_i^2}\right),
    \label{SingleSpeciesSolution}
\end{equation}
where $C_i$ is an undetermined constant.

\subsubsection{Two-Species Hydrodynamics Equations}
In the two species case,
the ground state density profile can be obtained from the TF approximation as
\begin{equation}
    \begin{pmatrix}
        g_{11} & g_{12}\\
        g_{12} & g_{22}
    \end{pmatrix}
    \begin{pmatrix}
        n_1\\
        n_2
    \end{pmatrix}
    =
    \begin{pmatrix}
        \mu_1\\
        \mu_2
    \end{pmatrix}
    +
    \frac{1}{2}\omega_0^2 r^2
    \begin{pmatrix}
        m_1\\
        m_2
    \end{pmatrix},
    \label{TF2Species}
\end{equation}
where $n_i$ is obtained by inverting the interaction matrix on the left.
To describe the excitation, the hydrodynamics equations that describe the BECs in terms of the densities and the velocity field $\mathbf{v}_i(\mathbf{r},t)$ are
\begin{equation}
    \begin{aligned}
            m_i\frac{\partial \mathbf{v}_i}{\partial t}=-\nabla&\left(\frac{1}{2}m_i(v_i^2+\omega_0^2r^2)+g_{ii}n_i+g_{12}n_j\right),
        \\
            &\frac{\partial n_i}{\partial t}+\nabla\cdot(n_i\mathbf{v}_i)=0.
    \end{aligned}
    \label{hydrodynamicsEqsFull}
\end{equation}
To analyze the excitations, we follow the single-species case and substitute $n_i(\mathbf{r},t)\mapsto n_i(r)+\delta n_i(\mathbf{r},t)$ and $\mathbf{v}_i(\mathbf{r},t)\mapsto 0+\delta \mathbf{v}_i(\mathbf{r},t)$ into \eq{hydrodynamicsEqsFull}. Up to first order in $\delta n_i$ and $\delta \mathbf{v}_i$, two equations in \eq{hydrodynamicsEqsFull} is combined into a single one, and substituting 
$\delta n_i(\mathbf{r},t)=\delta n_i(\mathbf{r}) e^{\mathbf{i}\omega t}$ yields 
\begin{equation}
    -m_i\omega^2\delta n_i=\nabla\cdot\left[n_{i}\nabla\left(  g_{ii}\delta n_i+g_{12}\delta n_j\right)\right],
    \label{CoreEquation}
\end{equation}
where the time-dependent phase is cancelled out and this equation determines the frequency spectrum.

\subsubsection{Perturbative analysis}

This section obtains a perturbative description for the breathing mode when $g_{12}\sim 0$ and $^{23}$Na encloses $^{87}$Rb ($R_1\gtrsim R_2$). We focus on $^{87}$Rb because only the overlapping region needed to be considered. 

We analyze the \eq{CoreEquation} by keeping up to first-order in $g_{12}$, with quantities being expanded as
\begin{equation}
    \begin{aligned}
        &n_i(r)=n^{(0)}_i(r)+g_{12}\Delta n_i(r),
        \\
        &\mu_i=\mu_i^{(0)}+g_{12}\Delta \mu_i,
        \\
        &\delta n_1(r,t)=s (\delta n^{(0)}_1(r)+g_{12}h_1(r))e^{-\mathbf{i}(\sqrt{5}\omega_0+g_{12}\omega^{\prime})t},
        \\
        &\delta n_2(r,t)=(\delta n^{(0)}_2(r)+g_{12}h_2(r))e^{-\mathbf{i}(\sqrt{5}\omega_0+g_{12} \omega^{\prime})t},
    \end{aligned}
    \label{ExpansionQuantity}
\end{equation}
where the superscript ``0'' 
is to denote the case when $g_{12}=0$, say, $n_i^{(0)}$ is the ground state distribution in the absence of inter-species interaction.
The first-order correction in frequency is $g_{12}\omega^{\prime}$, and $s=+1$ for the in-phase mode and $s=-1$ for the out-of-phase mode. To find $\omega^{\prime}$ and $h_i(r)$, we obtain the $\Delta n_i(r)$ through \eq{TF2Species} and substitute directly the expansion in \eq{ExpansionQuantity} into \eq{CoreEquation}, leading to a complicated equation. We can simplify the equation by rewriting $n^{(0)}_1\nabla \delta n^{(0)}_2$ and $n^{(0)}_2\nabla \delta n^{(0)}_1$ with the explicit expressions in \eqs{SingleSpeciesTF}{SingleSpeciesSolution}, and further postulating that the constants $C_i$ in \eq{SingleSpeciesSolution} are related by
\begin{equation}
    \frac{C_1}{C_2}=\frac{\frac{1}{2}\frac{m_1}{g_{11}}\omega_0^2 R_1^2}{\frac{1}{2}\frac{m_2}{g_{22}}\omega_0^2 R_2^2}=\frac{g_{22}m_1 R_1^2}{g_{11}m_2 R_2^2}.
\end{equation}
This means that the peak excitation amplitude is proportional to the peak density of the ground state distribution. Eventually, we cancel some terms using \eq{SingleSpeciesMain}, and obtain the eigenvalue equations
\begin{equation}
    \begin{aligned}
    -\sqrt{5}\omega_0 m_2 \delta n_2^{(0)}\left(2 \omega^{\prime}+(1 -s )\frac{m_1}{g_{11}m_2}\sqrt{5}\omega_0\right)
    =
    B+5 m_2\omega_0^2 h_2+\nabla\cdot[g_{22}n_2^{(0)}\nabla h_2]
    =
    5 m_2\omega_0^2  \tilde{h}_2+\nabla\cdot[g_{22}n_2^{(0)}\nabla \tilde{h}_2],
    \end{aligned}
    \label{ftilde}
\end{equation}
where $B$ is a constant
and $\tilde{h}_2\equiv h_2+B/(5\omega_0^2 m_2)$. There is a simple solution to this equation when $\tilde{h}_2$ has the same form as \eq{SingleSpeciesSolution}, such that $\nabla\cdot[g_{22}n_2^{(0)}\nabla \tilde{h}_2]=-5 m_2\omega_0^2 \tilde{h}_2$ and the term in parentheses in \eq{ftilde} equal to zero, giving rise to
\begin{equation}
    \omega^{\prime}=
    \begin{cases}
        0 &, \text{in-phase}
        \\
        -\sqrt{5}\omega_0\frac{m_1}{g_{11}m_2} &, \text{out-of-phase}.
    \end{cases}
    \label{}
\end{equation}
The in-phase mode frequency is again a constant $\sqrt{5}\omega_0$. The out-of-phase excitation frequency is
\begin{equation}
    \omega=\sqrt{5}\omega_0+g_{12}\omega^{\prime}=\sqrt{5}\omega_0 \left(1-\frac{g_{12}m_1}{g_{11}m_2}\right).
    \label{PerturbationResult}
\end{equation}
Such a solution of $h_2$ means that the inter-species interaction solely modulates the amplitude $\delta n_2^{(0)}$ and introduces a constant shift, hinting that $^{23}$Na acts as a background for $^{87}$Rb. 
To obtain $h_1$, one can flip the indices in \eq{ftilde} and substitute the determined $\omega^{\prime}$, then solve the resultant equation. For the in-phase mode, $h_1$ has a similar form as \eq{SingleSpeciesSolution} up to a constant. Hence both species retain the single-species nature.
For the out-of-phase mode, $h_1$does not have analytical form and is beyond our interest.



\subsection{ Simplified BdGEs in the Thin-shell Limit for $N_1 \ll N_2$}

\begin{figure}[t]
    \centering
    \includegraphics[width=0.6\linewidth]{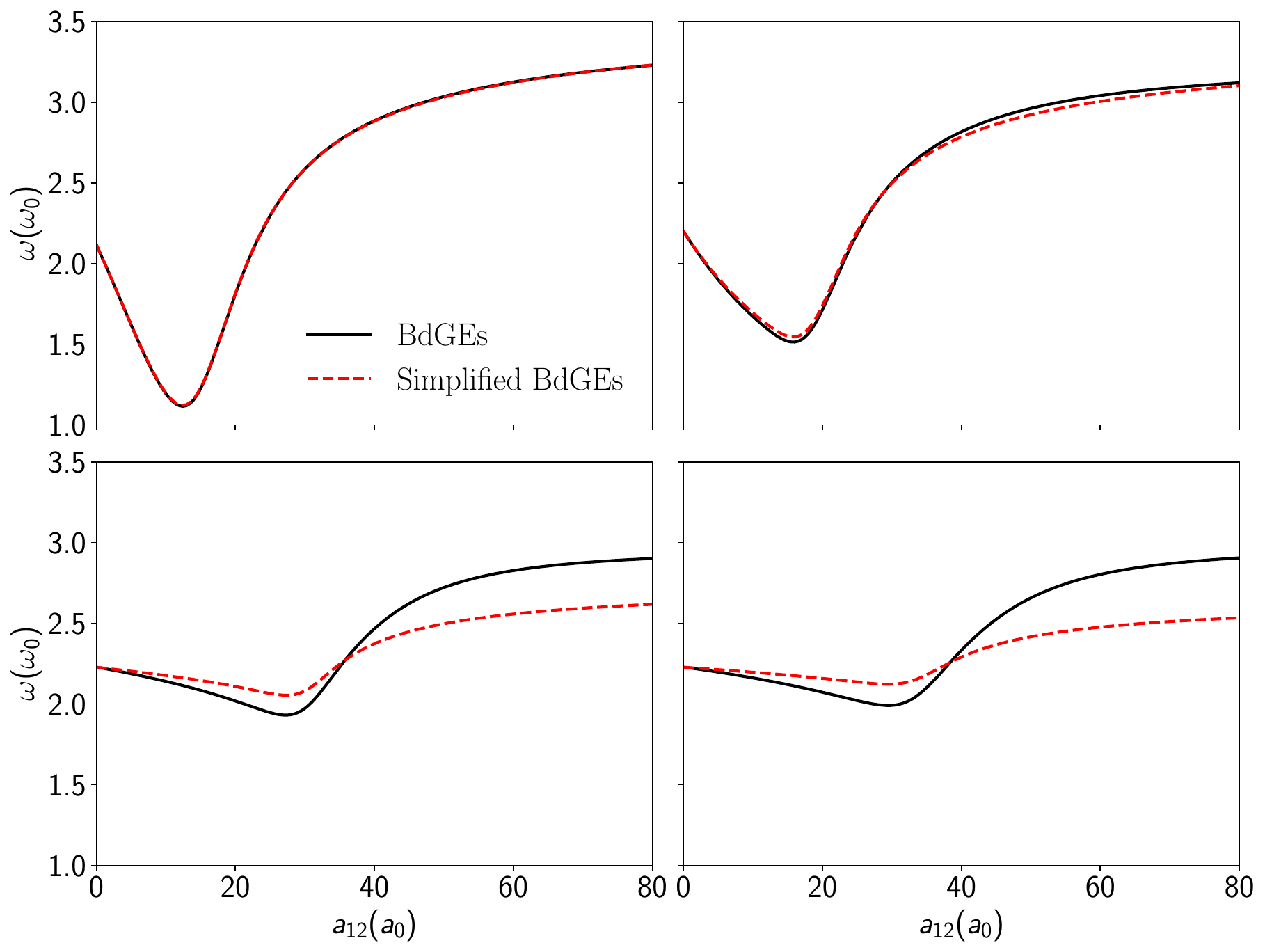}
    \caption{Comparison of the lowest out-of-phase monopole mode spectrum with full BdGEs (solid curves) and the simplified BdGEs (dashed curves) for several different atom numbers: 
    (a) $N_{1}=10^3$ and $N_{2}=10^5$.
    (b) $N_{1}=10^4$ and $N_{2}=10^5$.
    (c) $N_{1}=10^5$ and $N_{2}=10^5$.
    (d) $N_{1}=10^5$ and $N_{2}=7 \times 10^4$. 
    } 
    \label{SuppFig1}
\end{figure}

We focus on the case where $^{87}$Rb encloses $^{23}$Na ($R_2>R_1$). 
For weak $g_{12}$, 
inverting the interaction matrix in \eq{TF2Species} and expanding $g_{12}$ to first order yield the $^{23}$Na's density distribution
\begin{equation}
    n_{1}(r)=\frac{1}{2} \frac{m_1}{g_{11}}\tilde{\omega}_0^2 (\tilde{R}_{1}^2-r^2),
\end{equation}
where $\tilde{\omega}_0=\omega_0(1-g_{12}m_2/(2 g_{22}m_1))$ is the effective trapping frequency and $\tilde{R}_1>R_1$ is the TF radius under small $g_{12}$. The explicit form of $\tilde{R}_1$ can be found from the conservation of the number of particles, but it is not needed here.
The single-species BdGEs when $g_{12}=0$ and under TF approximation are \cite{Pethick2008}
\begin{equation}
    \begin{aligned}
        &\left( -\frac{\hbar^2}{2 m_1}\nabla^2 + \frac{1}{2}m_1\omega^2_0\left(R_1^2-r^2\right)\right)u(r) +\left(\frac{1}{2}m_1\omega^2_0\left(R_1^2-r^2\right)\right)v(r) =\hbar \omega u(r),
        \\
        &\left( -\frac{\hbar^2}{2 m_1}\nabla^2 + \frac{1}{2}m_1\omega^2_0\left(R_1^2-r^2\right)\right)v(r) +\left(\frac{1}{2}m_1\omega^2_0\left(R_1^2-r^2\right)\right)u(r)=-\hbar \omega v(r),
    \end{aligned}
    \label{BdGESingleSpecies}
\end{equation}
where $u(r)$ and $v(r)$ relates to the wavefunction as
$\psi(r,t)=\left(\psi^{(0)}(r)+u(r)e^{-i\omega t} +v^{*}(r)e^{i\omega t}\right)e^{-i\mu t}$. $\psi^{(0)}(r)$ is the ground state wavefunction.
Note that $\mu_1 - \frac{1}{2}m_1\omega^2_0r^2=g_{11}n_1^{(0)}$.

The two-species simplified BdGEs are obtained by neglecting the excitations of $^{87}$Rb. Denoting $\psi_1(r,t)=\left(\psi_1^{(0)}(r)+u(r)e^{-i\omega t} +v^{*}(r)e^{i\omega t}\right)e^{-i\mu_1 t}$, the simplified BdGEs are
\begin{equation}
    \begin{aligned}
        &\left( -\frac{\hbar^2}{2 m_1}\nabla^2+\frac{1}{2}m_1\omega^2_0 r^2 -\mu_1 +2 g_{11} n_1  + g_{12} n_2\right)u(r) +g_{11} n_1 v(r) =\hbar \omega u(r),
        \\
        &\left( -\frac{\hbar^2}{2 m_1}\nabla^2+\frac{1}{2}m_1\omega^2_0r^2 -\mu_1 +2 g_{11} n_1  + g_{12} n_2\right)v(r) +g_{11} n_1 u(r) =-\hbar \omega v(r).
    \end{aligned}
    \label{BdGEffective}
\end{equation}
In Fig.~\ref{SuppFig1}, we show the predicted out-of-phase spectra from numerically solving the full BdGEs and \eq{BdGEffective} for different $N_1$ and $N_2$ as a comparison. We see that the simplified BdGEs provide excellent descriptions for small $N_1$.
In TF approximation, since $\mu_1=\frac{1}{2}m_1\omega_0^2 r^2 +g_{11} n_1 + g_{12}n_2$ in \eq{TF2Species}, we find the terms inside the brackets in \eq{BdGEffective} can be simplified as
\begin{equation}
    \frac{1}{2}m_1\omega^2_0 r^2 -\mu_1 +2 g_{11} n_1  + g_{12} n_2 
    = g_{11} n_1
    =\frac{1}{2}m_1\tilde{\omega}_0^2 \left(\tilde{R}_1^2-r^2\right).
    \label{EffectiveOmega}
\end{equation}
Therefore \eq{BdGEffective} reduces to the form of \eq{BdGESingleSpecies} with effective trap frequency $\tilde{\omega}_0$. If $^{23}$Na is enclosed by $^{87}$Rb ($R_2>R_1$), the TF analysis for \eq{BdGEffective} is valid for the whole distribution, where we obtain
\begin{equation}
    \omega=\sqrt{5}\tilde{\omega}_0=\sqrt{5}\omega_0\left(1-\frac{g_{12}m_2}{2g_{22}m_1}\right).
    \label{EnclosedPerturb}
\end{equation}
On the other hand, if $^{23}$Na encloses $^{87}$Rb ($R_1>R_2$), we have to consider the $^{23}$Na density in the non-overlapping region. \eq{EffectiveOmega} is still true, but we have to consider the $^{87}$Rb's excitation with effective frequency $\omega_0\left(1-\frac{g_{12}m_1}{2g_{11}m_2}\right)$ higher than $^{23}$Na's and the non-overlapping region with the original frequencies $\omega_0$. Hence \eq{EnclosedPerturb} serves as a lower bound. Regardless of the cases, $\omega$ is lower than the single species value $\sqrt{5}\omega_0$.



\subsection{Variational Method}

\subsubsection{Before hollowing transition}

The Lagrangian for double species BECs is \cite{Pethick2008}
\begin{equation}
    \mathcal{L}=\int \left[\sum_i\left( \frac{\mathbf{i}\hbar}{2}\left(\psi_i^* \frac{\partial \psi_i}{\partial t}-\psi_i\frac{\partial \psi_i^*}{\partial t}\right) -\frac{\hbar^2}{2m_i}|\nabla \psi_i|^2 - V_i|\psi_i|^2 - \frac{g_{ii}}{2}|\psi_i|^4\right)  -g_{12}|\psi_1|^2|\psi_2|^2 \right] d^3r .
    \label{Lagrangian}
\end{equation}
By assuming different forms of ansatzes, we describe the dual-species BECs in different regimes. For weak $g_{12}$, the oscillations of two species resemble that in the single-species case where the width of the condensate is oscillating \cite{Pethick2008}, hence the ansatzes are set to be
\begin{equation}
    \psi_1(r,t)=\frac{\sqrt{N_1}}{\sigma_1(t)^{3/2}}f_1\left(\frac{r}{\sigma_1(t)}\right) e^{\mathbf{i}\beta_1(t) m_1 r^2/(2\hbar)}
    , \qquad
    \psi_2(r,t)=\frac{\sqrt{N_2}}{\sigma_2(t)^{3/2}}f_2\left(\frac{r}{\sigma_2(t)}\right) e^{\mathbf{i}\beta_2(t) m_2 r^2/(2\hbar)},
    \label{Variation0_Ansatez}
\end{equation}
where $\sigma_1(t)$, $\sigma_2(t)$, $\beta_1(t)$, $\beta_2(t)$ are the variational parameters, and the functions $f_1(x)$ and $f_2(x)$ do not need to be known explicitly. 
$r$ is the radial distance from the harmonic trap center.
The length $\sigma_i(t)$ is the width of the distribution, and it is equal to a stationary value $\sigma_i^{(0)}$ in the ground state.
The ground state density profile defines $\sigma_i^{(0)}$ once the function $f_i(x)$ is chosen explicitly. Equivalently, \eq{Variation0_Ansatez} represents the ground state distribution when $\sigma_i(t)=\sigma_i^{(0)}$ (and $\beta_i(t)=0$). During the excitation, $\sigma_i(t)$ oscillates periodically around the equilibrium value with a small amplitude deviation $\tilde{\sigma}_i(t) \equiv \sigma_i(t) - \sigma_i^{(0)}$.
The form of the phase $\beta_i(t) m_i r^2/(2\hbar)$ represents a velocity field $ \beta_i(t) r \hat{r} $, where $\hat{r}$ is the radial unit vector. This choice of the phase characterizes the species' motion to be its width $\sigma_i(t)$.
In the following, $\sigma_i$ and $\sigma_i(t)$ are used synonymously, and similarly for other parameters.
Putting the ansatzes into the Lagrangian \eq{Lagrangian}, we obtain
\begin{equation}
    \mathcal{L}=-\left\{ \left[\frac{1}{2}m_1N_1\sigma_1^2c_{\text{tr,}1}(\dot{\beta_1}+\beta_1^2) + U_1(\sigma_1)\right]+\left[\frac{1}{2}m_2N_2\sigma_2^2c_{\text{tr,}2}(\dot{\beta_2}+\beta_2^2) + U_2(\sigma_2)\right]  +U_{12}\right\},
\end{equation}
where
\begin{equation}
    \begin{aligned}
        &U_i(\sigma_i)=\frac{c_{\text{zp,}i}}{\sigma_i^2}+\frac{1}{2}m_iN_i\omega_0^2\sigma_i^2c_{\text{tr,}i}+\frac{c_{\text{int,}i}}{\sigma_i^3} \Rightarrow E_{\text{zp,}i} + E_{\text{tr,}i}+E_{\text{int,}i},
        \\
        &U_{12}(\sigma_1,\sigma_2)=g_{12}\frac{N_1N_2}{\sigma_1^3\sigma_2^3}\int f_1^2\left(\frac{r}{\sigma_1}\right)f_2^2\left(\frac{r}{\sigma_2}\right)d^3r \Rightarrow E_{\text{int,}12}.
    \end{aligned}
\end{equation}
The `$\Rightarrow$' denotes what the expression reduces to when $\sigma_i$ takes the equilibrium value in the ground state. $E_{\text{zp,}i}=\int \hbar^2/(2m_i)\nabla^2 n_i d^3r$, $E_{\text{tr,}i}=\int (1/2) m_i\omega_0^2 r^2 n_i d^3r$, $E_{\text{int,}i}=g_{ii}\int n_i^2 d^3r$, and $E_{\text{int,}12}=g_{12}\int n_1 n_2 d^3r$
are respectively the zero-point (kinetic), potential, interaction energies for species $i$, and the interspecies interaction energy. $n_i=n_i(r)=|\psi_i(r)|^2$. The `$c$' terms like $c_{\text{zp}1}$ are constants and need not to be determined explicitly. From Lagrange equations, we find $\beta_i=\frac{\dot{\sigma}_i}{\sigma_i}$, and that
 for $\sigma_i$ yields
\begin{equation}
    \begin{aligned}
        m_i N_i c_{\text{tr}i} \ddot{\sigma}_i&=-\frac{\partial }{\partial \sigma_i}(U_i+U_{12})
        \\
        &=
        -\frac{1}{\sigma_i}
        \left[ -2 \frac{c_{\text{zp,i}}}{\sigma_i^2} + m_i N_i \omega_0^2 \sigma_i^2 c_{\text{tr,i}} -3 \frac{c_{\text{int,i}}}{\sigma_i^3}    
        -
        g_{12}\frac{N_1N_2}{\sigma_1^3\sigma_2^3}\left(3\int f_1^2 f_2^2 d^3r+\int r \frac{\partial f_i^2}{\partial r} f_j^2 d^3r\right)
        \right].
    \end{aligned} 
    \label{Variation0_1order}
\end{equation}
Given the system is in the ground state where $\sigma_i=\sigma_i^{(0)}$ and $\ddot{\sigma}_i=0$, it leads to the condition
\begin{equation}
    -2 E_{\text{zp,i}} +2 E_{\text{tr,i}} -3 E_{\text{int,i}}    
    -
    3 E_{\text{int,12}}
    -g_{12}\int r \frac{\partial n_i}{\partial r} n_j d^3r =0,
\end{equation}
This condition is numerically verified with the solution from GPE. Directly adding of the above equations for $(i,j)=(1,2)$ and $(i,j)=(2,1)$ and using integration by part produce
\begin{equation}
    -2 (E_{\text{zp,1}}+E_{\text{zp,2}}) +2 (E_{\text{tr,1}}+E_{\text{tr,2}}) -3 (E_{\text{int,1}}+ E_{\text{int,12}} +E_{\text{int,2}})    
    =0,
    \label{Virial0}
\end{equation}
which resembles the virial condition for the single-species case.
Then, we assume the oscillating amplitude of $\sigma_i$ is much smaller than their respective equilibrium values and expand the force term in \eq{Variation0_1order} up to linear order in $\tilde{\sigma}_i$. The equation of motion is
\begin{equation}
    -m_i N_i c_{\text{tr}i} \ddot{\tilde{\sigma}}_i=
    \tilde{\sigma}_i\frac{\partial^2 }{\partial \sigma_i^2}\left( U_i+U_{12} \right)\bigg|_{\sigma_i=\sigma_i^{(0)}}
    + \tilde{\sigma}_j\frac{\partial^2 U_{12}}{\partial \sigma_1 \partial \sigma_2}\bigg|_{\sigma_i=\sigma_i^{(0)}}.
    \label{Variation0_MainEqs}
\end{equation}
We note that we can symbolically rewrite \eq{Variation0_MainEqs} as
\begin{equation}
    -
    \begin{pmatrix}
        \alpha_1 & 0
        \\
        0 & \alpha_2
    \end{pmatrix}
    \begin{pmatrix}
        \ddot{\tilde{\sigma}}_1
        \\
        \ddot{\tilde{\sigma}}_2
    \end{pmatrix}
    = 
    \begin{pmatrix}
        \chi_1 & \delta
        \\
        \delta & \chi_2  
    \end{pmatrix}
    \begin{pmatrix}
        \tilde{\sigma}_1
        \\
        \tilde{\sigma}_2
    \end{pmatrix},
\end{equation}
where $\alpha_i \equiv m_i N_i c_{\text{tr},i}$. Explicitly, the symbols represent
\begin{equation}
    \begin{aligned}
        \left(\sigma_i^{(0)}\right)^2 \chi_i &\equiv
        \sigma_i^2 \frac{\partial^2 }{\partial \sigma_i^2}\left( U_i+U_{12} \right)\bigg|_{\text{equil}}
        =
        6 E_{\text{zp,}i} + 2 E_{\text{tr,}i}+ 12 E_{\text{int,}i} + 12 E_{\text{int,}12} + 8 g_{12} \int r \frac{\partial n_i}{\partial r} n_j d^3r +
        g_{12} \int r^2 \frac{\partial^2 n_i}{\partial r^2} n_j d^3r,
        \\
        \sigma_1^{(0)} \sigma_2^{(0)} \delta &\equiv
        \sigma_1 \sigma_2 \frac{\partial^2 U_{12}}{\partial \sigma_1 \partial \sigma_2}\bigg|_{\text{equil}} = g_{12}\int r^2 \frac{\partial n_1}{\partial r}\frac{\partial n_2}{\partial r} d^3r.
    \end{aligned}
    \label{Variation0_SecondOrder}
\end{equation}
Since all the coefficients are known and \eq{Variation0_MainEqs} just represents the coupled harmonic oscillators, we can obtain the excitation frequencies in terms of the ground state density distributions.  The eigenfrequencies are
\begin{equation}
    \omega^2= \frac{\chi_1}{2 \alpha_1} + \frac{\chi_2}{2 \alpha_2} \pm \frac{1}{2}\sqrt{4\frac{\delta^2}{\alpha_1 \alpha_2} + \left(\frac{\chi_1}{\alpha_1} - \frac{\chi_2}{\alpha_2}\right)^2},
    \label{Variation0_Result}
\end{equation}
where see that the frequencies are unchanged if $\alpha_i \rightarrow \alpha_i \left(\sigma_i^{(0)}\right)^2 = 2 E_{\text{tr,}i}$, $\chi_i \rightarrow \chi_i \left(\sigma_i^{(0)}\right)^2$, and $\delta \rightarrow \sigma_1^{(0)} \sigma_2^{(0)} \delta$. Therefore, we can use expression in \eq{Variation0_SecondOrder} without the explicit determination of $\sigma_i^{(0)}$. The frequencies obtained by \eq{Variation0_Result} using the numerical ground state distributions from GPE are consistent with the results from BdGEs for small $a_{12}$.

\subsubsection{After hollowing transition}

\begin{figure}[t]
    \centering
    \includegraphics[width=0.5\linewidth]{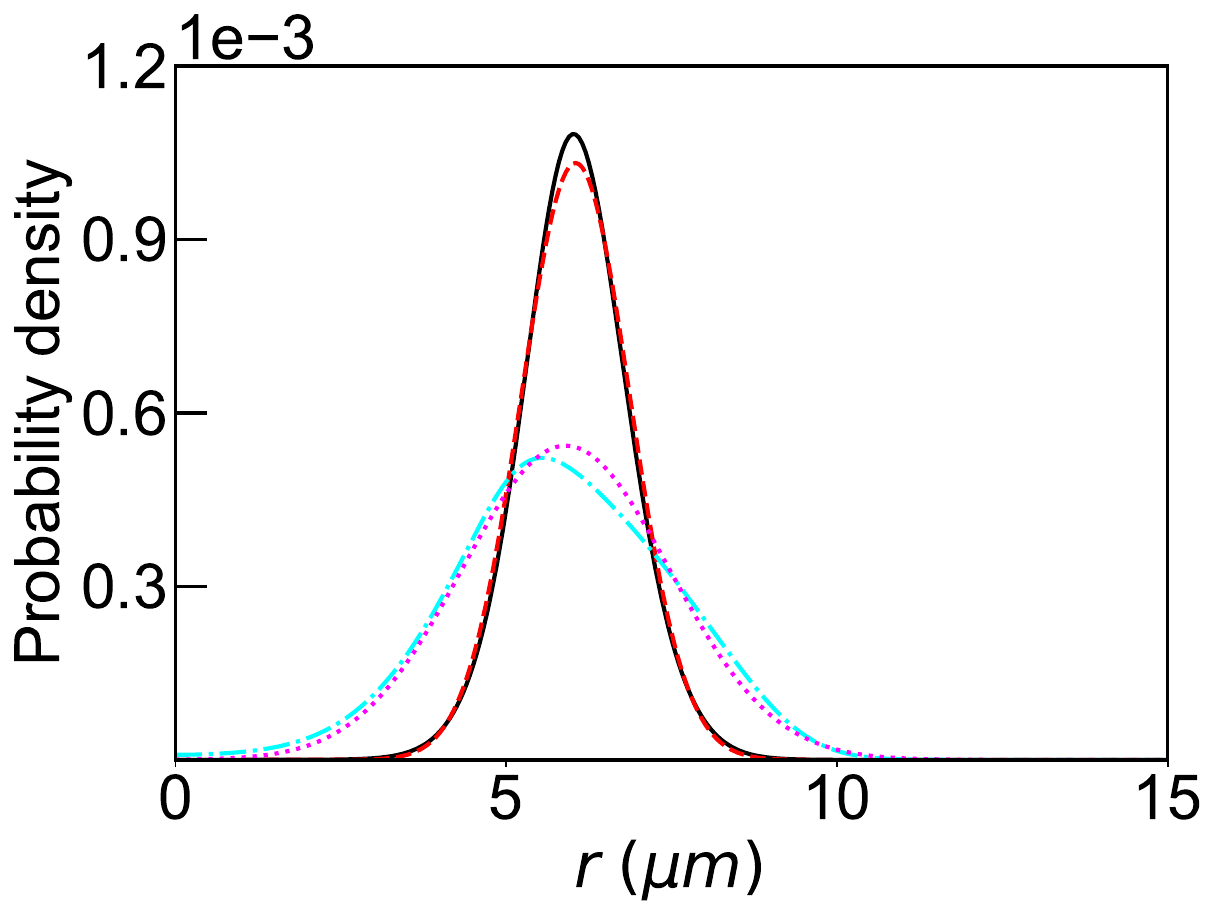}
    \caption{ 
        $a_{12}=36a_0$, which is beyond the hollowing transition points, for all curves.
        Black line: BdGEs, $N_1=10^3$, $N_2=10^5$. Red dashed line: Gaussian ansatzes, $N_1=10^3$, $N_2=10^5$.
        Aqua dashed-dotted: BdGEs, $N_1=10^5$, $N_2=7\times 10^4$.
        Magenta dotted: Gaussian ansatzes, $N_1=10^5$, $N_2=7\times 10^4$.
    } 
    \label{SuppFig2}
\end{figure}

Here, we describe the breathing modes when $^{23}$Na has already formed a shell. We employ ansatzes similar to that in \cite{Lannert2007}, 
with the coupling between the width and the radial center motion of the shell species included. 
Explicitly, the ansatzes are
\begin{equation}
    \begin{aligned}
        &\psi_1(r,t)=\frac{\sqrt{N_1}}{\sqrt{2 \pi^{3/2} \sigma_1 (2 r_c^2+\sigma_1^2)}} e^{-\frac{1}{2}\left(\frac{r-r_c}{\sigma_1}\right)^2}
        e^{\mathbf{i}\beta_0 m_1 r/\hbar  +  \mathbf{i}\beta_1 m_1 (r-r_c)^2/(2\hbar)},
        \\
        &\psi_2(r,t)=\frac{\sqrt{N_2}}{\sigma_2^{3/2}}f_2\left(\frac{r}{\sigma_2}\right) e^{\mathbf{i}\beta_2 m_2 r^2/(2\hbar)},
    \end{aligned}
    \label{Variation3_ansatzes}
\end{equation}
where $r_c(t)$, $\sigma_1(t)$, $\sigma_2(t)$, $\beta_0(t)$, $\beta_1(t)$, $\beta_2(t)$ are variational parameters.
$r_c(t)$ is a new parameter characterizing the radial center of the $^{23}$Na's distribution. The comparison with the Sodium density distribution from GPE and this ansatz is shown in Fig.~\ref{SuppFig2}.
The velocity field for $^{23}$Na is $\beta_0(t)\hat{r} + \beta_1(t)(r-r_c)\hat{r}$, such that $\beta_0$ characterizes the radial center's motion.
Generally, $r_c$ and $\sigma_1$ have complicated coupling. To simplify the situation we assume $^{23}$Na forms a thin shell, meaning that \cite{Lannert2007}
\begin{equation}
    r_c \gg \sigma_1,
\end{equation}
such $\sigma_1/r_c$ is an expansion parameter. Indeed, this expansion causes the result in this section to be only valid after the hollowing transition.
As a technical remark, we approximate $\text{Erf}(r_c/\sigma_1) \approx 1$ and $e^{-(r_c/\sigma_1)^2}\approx 0$ due to their rapid convergence, here $\text{Erf}(x)$ is the error function.
Furthermore, we assume the lengths to oscillate in very small amplitudes to keep harmonic motions, such that the ordering is
\begin{equation}
    \left(\frac{\tilde{\sigma}_1}{\sigma_1}  \sim  \frac{\tilde{r}_c}{r_c} \right) \ll \frac{\sigma_1^2}{r_c^2}, \qquad  \frac{\tilde{\sigma}_1}{\tilde{r}_c} \sim \frac{\sigma_1}{r_c},
\end{equation}
where $\tilde{\sigma}_1$ is the deviation of $\sigma_1$ from its equilibrium value.
It means that, up to second order in $\sigma_1/r_c$, terms like $\frac{\dot{\sigma}_1}{r_c}\dot{\sigma}_1=\frac{\dot{\tilde{\sigma}}_1}{r_c}\dot{\tilde{\sigma}}_1$ are discarded. $\tilde{\sigma}_1/\sigma_1  \sim  \tilde{r}_c/r_c$ because $\tilde{\sigma}_1/\sigma_1$ needs to be comparable to $\tilde{r}_c/r_c$ for the strong coupling to occur for a thick shell. In the thin shell limit ($\sigma_1\rightarrow 0$), the width is frozen so $\tilde{\sigma}_1\rightarrow 0$.

The Lagrangian is $\mathcal{L}=\mathcal{L}_1+\mathcal{L}_2 - U_{12}$, up to second order we have
\begin{equation}
    \begin{aligned}
    &\mathcal{L}_1=
    -m_1\left[
     \frac{1}{2} \beta_0^2
     + \frac{\sigma_1}{r_c} (\beta_0 \beta_1- \beta_1 \dot{r}_c +\dot{\beta}_0) \sigma_1 
     +\frac{1}{4} \left(1 + \frac{\sigma_1^2}{r_c^2}\right)\beta_1^2 \sigma_1^2 
     + \dot{\beta}_0 r_c
     +\frac{1}{4} \dot{\beta}_1 \sigma_1^2 
    \right]-U_1(r_c,\sigma_1),
    \\
    &\mathcal{L}_2=-\left[\frac{1}{2}m_2N_2\sigma_2^2c_{\text{tr,}2}(\dot{\beta_2}+\beta_2^2) + U_2(\sigma_2)\right].
    \end{aligned}
\end{equation}
In practical calculation, for terms involving $\beta_i$ and its derivative, we keep up to fourth order in $\sigma_1/r_c$. It is because $\dot{r}_c/\dot{\sigma}_1\sim r_c/\sigma_1$, term like $\frac{\sigma_1^3}{r_c^3}\frac{\dot{r}_c}{\dot{\sigma}_1}$ is in fact a second order term. 
The explicit expressions are
\begin{equation}
    \begin{aligned}
        &U_1
        =
        N_1\left[\frac{\hbar^2}{4 m_1 \sigma_1^2}\left(1+\frac{\sigma_1^2}{r_c^2}\right)
        +
        \frac{1}{2}m_1 \omega_0^2 r_c^2\left(1 + \frac{5}{2}\frac{\sigma_1^2}{r_c^2} \right)
        +
        \frac{g_{11} N_1}{8\sqrt{2}\pi^{3/2} \sigma_1 r_c^2}\left(1+\frac{3}{4}\frac{\sigma_1^2}{r_c^2}\right)\right]
        \Rightarrow
        E_{\text{zp,}1} + E_{\text{tr,}1}+E_{\text{int,}1},
        \\
        &U_2
        =
        \frac{c_{\text{zp,}2}}{\sigma_2^2}+\frac{1}{2}m_2N_2\omega_0^2\sigma_2^2c_{\text{tr,}2}+\frac{c_{\text{int,}2}}{\sigma_2^3} \Rightarrow E_{\text{zp,}2} + E_{\text{tr,}2}+E_{\text{int,}2},
        \\
        &U_{12}=\frac{g_{12}N_1N_2}{2 \pi^{3/2} \sigma_1 (2 r_c^2+\sigma_1^2)\sigma_2^3}\int f_1^2\left(\frac{r-r_c}{\sigma_1}\right)f_2^2\left(\frac{r}{\sigma_2}\right)d^3r \Rightarrow E_{\text{int,}12},
    \end{aligned}
\end{equation}
where $f_1(x)=e^{-x^2/2}$.
We can determine, without fitting, the explicit value of $r_c^{(0)}$ and $\sigma_1^{(0)}$, which are the equilibrium values, from $E_{\text{zp,}1}$ and $E_{\text{tr,}1}$. For the velocity field parameters, we find
\begin{equation}
    \begin{aligned}
        \beta_0=\left(1-\frac{\sigma_1^2}{r_c^2}\right)\dot{r}_c + \frac{\sigma_1}{r_c} \dot{\sigma}_1,
        \qquad
        \beta_1=\left(1-\frac{\sigma_1^2}{r_c^2}\right) \frac{\dot{\sigma}_1}{\sigma_1} + \frac{\sigma_1^2}{r_c^2}\frac{\dot{r}_c}{r_c},
    \end{aligned}
\end{equation}
which exhibit the coupling between the width and the radial center motions. In the following, we rewrite the equillibrium values $r_c^{(0)}$, $\sigma_1^{(0)}$, and $\sigma_2^{(0)}$ as $r_c$, $\sigma_1$, and $\sigma_2$ for brevity. The equilibrium conditions are
\begin{equation}
    \begin{aligned}
        &
        -N_{1}\frac{\hbar^2}{2 m_1 r_c^2} + 2E_{\text{tr,}1}-N_1\frac{5}{2}m_1 \omega_0^2 \sigma_1^2 - 2E_{\text{int,}1} + \frac{3 g_{11} N_{1}^2 \sigma_1 }{16\sqrt{2}\pi^{3/2}r_c^4}-\left(2-\frac{\sigma_1^2}{r_c^2}\right)E_{\text{int,}12}-g_{12}r_c\int\frac{\partial n_1}{\partial r} n_2 d^3r=0,
        \\
        &
        -2E_{\text{kin,}1}+N_1\frac{\hbar^2}{2m_1 r_c^2}+N_1\frac{5}{2}m_1\omega_0^2 \sigma_1^2 -E_{\text{int,}1}-\frac{3 g_{11} N_{1}^2 \sigma_1}{16\sqrt{2}\pi^{3/2}r_c^4} - \left(1+\frac{\sigma_1^2}{r_c^2}\right)E_{\text{int,}12} - g_{12}\int\left(r-r_c\right)\frac{\partial n_1}{\partial r} n_2 d^3r =0,
    \end{aligned}
\end{equation}
which are numerically correct for large $a_{12}$. The equations of motion are in the form of a generalized eigenvalue problem
\begin{equation}
    -
    \begin{pmatrix}
        \alpha_0 & \gamma_{01} & 0
        \\
        \gamma_{10} & \alpha_1 & 0
        \\
        0 & 0 & \alpha_2
    \end{pmatrix}
    \begin{pmatrix}
        \ddot{\tilde{r}}_c
        \\
        \ddot{\tilde{\sigma}}_1
        \\
        \ddot{\tilde{\sigma}}_2
    \end{pmatrix}
    =
    \begin{pmatrix}
        \chi_0 & \delta_{01} & \delta_{02}
        \\
        \delta_{01} & \chi_1 & \delta_{12}
        \\
        \delta_{02} & \delta_{12} & \chi_2
    \end{pmatrix}
    \begin{pmatrix}
        \tilde{r}_c
        \\
        \tilde{\sigma}_1
        \\
        \tilde{\sigma}_2
    \end{pmatrix}.
    \label{Variation3_MatrixEOM}
\end{equation}
The details for the matrix on the left-hand-side are
\begin{equation}
    \begin{aligned}
    &r_c^2\alpha_0=\left(1-2\frac{\sigma_1^2}{r_c^2}\right)2E_{\text{tr,}1}-N_1\frac{5}{2}m_1\omega_0^2\sigma_1^2
    ,\qquad 
    \sigma_1^2\alpha_1=N_1\frac{1}{2}m_1\omega_0^2\sigma_1^2
    ,\qquad 
    \sigma_2^2\alpha_2=2E_{\text{tr,}2},
    \\
    &r_c \sigma_1\gamma_{01}=N_1m_1\omega_0^2\sigma_1^2
    ,\qquad
    r_c \sigma_1\gamma_{02}=2 N_1m_1\omega_0^2\sigma_1^2.
    \end{aligned}
\end{equation}
The details for the matrix on the right-hand-side are
\begin{equation}
    \begin{aligned}
        r_c^2\chi_0=&\,N_1\frac{3\hbar^2}{2m_1 r_c^2} +2 E_{\text{tr,}1} -N_1\frac{5}{2}m_1\omega_0^2 \sigma_1^2 + 6 E_{\text{int,}1}-\frac{21 g_{11} N_1^2 \sigma_1}{16\sqrt{2}\pi^{3/2}r_c^4}+\left(6-7\frac{\sigma_1^2}{r_c^2}\right)E_{\text{int,}12} ,
        \\
        &+ 
        \left(4-2\frac{\sigma_1^2}{r_c^2}\right)g_{12}r_c\int\frac{\partial n_1}{\partial r} n_2 d^3r +
        g_{12}r_c^2\int\frac{\partial^2 n_1}{\partial r^2} n_2 d^3r ,
        \\
        \sigma_1^2\chi_1=&\,6 E_{\text{kin,}1}-N_1\frac{3\hbar^2}{2m_1r_c^2}+N_1\frac{5}{2}m_1\omega_0^2\sigma_1^2 + 2E_{\text{int,}1}+\frac{3g_{11}N_1^2\sigma_1}{16\sqrt{2}\pi^{3/2}r_c^4} + \left(2+\frac{\sigma_1^2}{r_c^2}\right)E_{\text{int,}12} ,
        \\
        &+ 
        \left(4+2\frac{\sigma_1^2}{r_c^2}\right)g_{12}\int\left(r-r_c\right)\frac{\partial n_1}{\partial r} n_2 d^3r +
        g_{12}\int\left(r-r_c\right)^2\frac{\partial^2 n_1}{\partial r^2} n_2 d^3r,
        \\
        \sigma_2^2\chi_2=&\,6E_{\text{kin,}2}+2E_{\text{tr,}2}+12E_{\text{int,}2}+12E_{\text{int,}12}+ g_{12}\int r n_1 \frac{\partial n_2}{\partial r}d^3r + g_{12}\int r^2 n_1 \frac{\partial^2 n_2}{\partial r^2}d^3r,
        \\
        r_c\sigma_1\delta_{01}=&\,2E_{\text{int,}1}+\frac{9g_{11}N_1^2\sigma_1}{16\sqrt{2}\pi^{3/2}r_c^4}+\left(2+3\frac{\sigma_1^2}{r_c^2}\right)E_{\text{int,}12}
        +
        \left(2+\frac{\sigma_1^2}{r_c^2}\right)g_{12}r_c\int\frac{\partial n_1}{\partial r} n_2 d^3r,
        \\
        &+
        \left(2-\frac{\sigma_1^2}{r_c^2}\right)g_{12}\int\left(r-r_c\right)\frac{\partial n_1}{\partial r} n_2 d^3r
        +
        g_{12}r_c\int\left(r-r_c\right)\frac{\partial^2 n_1}{\partial r^2} n_2 d^3r,
        \\
        r_c\sigma_2\delta_{02}=&\,\left(6-3\frac{\sigma_1^2}{r_c^2}\right)
        +
        3 g_{12}r_c\int\frac{\partial n_1}{\partial r} n_2 d^3r
        +
        \left(2-\frac{\sigma_1^2}{r_c^2}\right)g_{12}\int r n_1 \frac{\partial n_2}{\partial r}d^3r
        +
        g_{12}r_c\int r \frac{\partial n_2}{\partial r} \frac{\partial n_2}{\partial r}d^3r,
        \\
        \sigma_1 \sigma_2\delta_{12}=&\,
        3\left(1+\frac{\sigma_1^2}{r_c^2}\right)E_{\text{int,}12}
        +
        3g_{12}\int\left(r-r_c\right)\frac{\partial n_1}{\partial r} n_2 d^3r
        +
        \left(1+\frac{\sigma_1^2}{r_c^2}\right)g_{12}\int r n_1 \frac{\partial n_2}{\partial r}d^3r,
        \\
        &+
        g_{12}\int r\left(r-r_c\right) \frac{\partial n_1}{\partial r} \frac{\partial n_2}{\partial r}d^3r.
    \end{aligned}
\end{equation}

With the above explicit form, the excitation frequencies are obtained by diagonalizing the \eq{Variation3_MatrixEOM}. As a technical remark, we use the Gaussian ansatz for $n_1$ instead of the GPE solution in evaluating terms like $\int n_1 n_2 d^3r$ for self-consistency. As the $^{23}$Na shell gets thinner with decreasing $N_1$ ,the results from the variational method converge to that from BdGEs. However, the variational method cannot capture the trend for very large $a_{12}$ because our ansatz for $^{23}$Na in \eq{Variation3_ansatzes} does not capture the skewness of the wavefunction. The skewness is becoming important as $g_{12}$ increases, as we can see from the effective potential $V_{\text{eff}}(r)=\frac{1}{2}m_1\omega_0^2 r^2 + g_{12}n_2(r)$. As $g_{12}$ increases, $V_{\text{eff}}(r)$ turns into a hard wall and a harmonic trap potential, and the actual wavefunction becomes more skew-symmetric. A symmetric Gaussian function
either allows a penetration into the hard wall or shifts the radial center to the wrong location. 

\begin{figure}[h]
    \centering
    \includegraphics[width=0.6\linewidth]{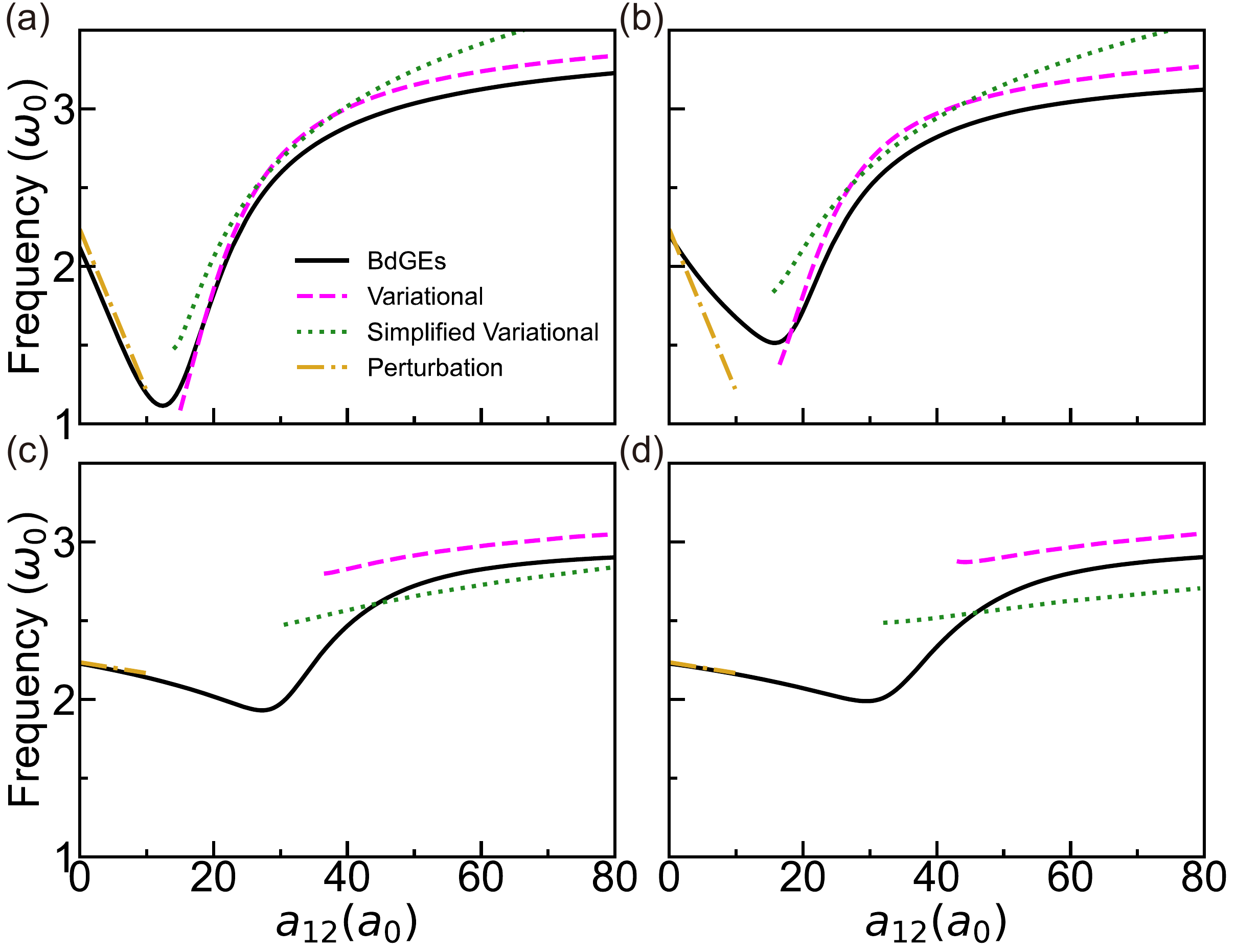}
    \caption{The theoretic prediction for the out-of-phase mode excitation frequency vs $a_{12}$ for different $N_{1}$ and $N_{2}$. Results from the BdGEs, variational method, simplified variational method (only look at $r_c(t)$), and perturbation are shown. The perturbation lines for (a) and (b) are $\sqrt{5}\omega_0\left(1-g_{12}m_{2}/(2 g_{22}m_{1})\right)$,
    while for (c) and (d) are $\sqrt{5}\omega_0\left(1-g_{12}m_{1}/(g_{11}m_{2})\right)$.
    (a) $N_{1}=10^3$ and $N_{2}=10^5$. At $a_{12}=0a_0$, the excitation frequency is not $\sqrt{5}\omega_0$ because $N_{1}$ is not large enough.
    (b) $N_{1}=10^4$ and $N_{2}=10^5$.
    (c) $N_{1}=10^5$ and $N_{2}=10^5$.
    (d) $N_{1}=10^5$ and $N_{2}=7 \times 10^4$. 
    } 
    \label{EndNoteFig}
\end{figure}

In Fig.~\ref{EndNoteFig}, we show the results from the full and simplified variational methods (only $r_c(t)$ in motion) with the ground state distributions obtained from GPE for different particle numbers of the species. As expected, the results of the variational method converge to those of the BdGEs as the $^{23}$Na shell becomes thinner (from (d) to (a)). The simplified one also captures the frequency trend. The convergence of the variational method can be improved by taking into account the skewness of the $^{23}$Na density profile, but the mathematical difficulty increases.